\documentclass[11pt]{article}

\usepackage[T1]{fontenc}
\usepackage[utf8]{inputenc}
\usepackage{mathptmx}
\usepackage[a4paper,margin=2.3cm]{geometry}
\usepackage{amsmath,amssymb,amsthm}
\usepackage{graphicx}
\usepackage{booktabs,multirow}
\usepackage{caption}
\usepackage{subcaption}
\usepackage{authblk}
\usepackage{microtype}
\usepackage{algorithm}
\usepackage[noend]{algpseudocode}
\usepackage[numbers,sort&compress]{natbib}
\usepackage[dvipsnames]{xcolor}
\definecolor{linkcol}{RGB}{20,70,120}
\usepackage[colorlinks=true,allcolors=linkcol]{hyperref}
 \usepackage[title,titletoc]{appendix}
\theoremstyle{plain}
\newtheorem{proposition}{Proposition}[section]

\theoremstyle{definition}

\newcommand{\Uobs}{U^{\mathrm{obs}}}
\newcommand{\RelErr}{\mathrm{RelErr}}
\newcommand{\Efun}{\mathcal{E}}
\newcommand{\R}{\mathbb{R}}

\title{\bfseries A scale-balanced variational physics-informed neural network for the inverse SEIR reaction-diffusion problem}

\author[1]{Soumyajit Saha\thanks{Email: \href{mailto:soumyajit.saha@unistra.fr}{soumyajit.saha@unistra.fr}}}
\affil[1]{\small Institut de Recherche Math\'ematique Avanc\'ee (IRMA), Universit\'e de Strasbourg, 67084 Strasbourg, France}
\author[2]{Aniket Banerjee\thanks{Email: \href{mailto:aniket.banerjee@sorbonne-universite.fr}{aniket.banerjee@sorbonne-universite.fr}}}
\affil[2]{\small Laboratoire Jacques-Louis Lions, Sorbonne Universit\'e, CNRS, 75005 Paris, France}
\date{}

\begin{document}
\maketitle

\begin{abstract}
\noindent
Spatial epidemic models add a diffusion term to the SEIR equations to describe how an outbreak spreads across a region, where its coefficient, the diffusivity $\lambda$, sets the speed of that spread. We ask how well $\lambda$ and the transmission, progression and recovery rates can be recovered from noisy space-time data. The answer is uneven: the reaction rates are recovered correctly for clean data but $\lambda$ is stubbornly hard, for two reasons we make precise. The compartments differ in size by orders of magnitude for a typical disease dynamics model, so an ordinary least-squares fit spends $99.94\%$ of its effort on the susceptible population, and $\lambda$ multiplies the Laplacian, so it is fixed by the \emph{curvature} of the fields rather than their values, and any data fitting method can match the data closely while still getting that curvature wrong. The proposed scale-balanced variational physics-informed neural network (sb-VPINN) for the SEIR reaction-diffusion problem removes the first obstacle by rescaling each compartment, and the second by writing the dynamics in weak form, that is, integrating the diffusion term by parts against smooth test functions, so that only first derivatives are ever taken and the boundary terms vanish. Across five seeds, the weak formulation reduces the mean clean-data diffusion error from $50.35\%$ to $13.72\%$ in one dimension. Across all six noise levels, its mean reaction-parameter errors remain below $1.3\%$. Improved diffusion recovery relative to the scale-balanced strong form also persists on the two-dimensional square.
\end{abstract}


% =============================================================================
\section{Introduction}
% =============================================================================

Compartmental models have been a central object of mathematical epidemiology since Kermack and McKendrick \citep{kermack1927}, and the susceptible-exposed-infectious-recovered (SEIR) family remains a standard description of outbreak dynamics \cite{hethcote2000,vandendriessche2002}. In such models spatial spread is key, the well-mixed hypothesis\footnote{The well-mixed hypothesis assumes that the population is spatially homogeneous and that individuals mix uniformly, irrespective of their location.} \cite{bansal2007} fails and the compartments become spatially resolved densities coupled by diffusion. Such reaction-diffusion descriptions have a long history in modelling geographic invasion \cite{murray2003,riley2007} and were used at scale during the COVID-19 pandemic \cite{viguerie2021}. In such a model, the reaction terms encode transmission, incubation, recovery and demography, while the diffusion term encodes spatial mobility. Its diffusion coefficient $\lambda$ fixes the scale over which the epidemic invades new territory, and is thus a quantity of direct epidemiological interest rather than a nuisance parameter. The reaction terms are nonlinear and can support delicate dynamical behaviour like spatial population models, which exhibit finite-time blow-up dynamics, depending on the interaction structure \cite{parshad2023}. Thus, identifying their coefficients from data is not a routine process.

Recovering the governing coefficients from observed space-time data is a parameter inverse problem for a nonlinear parabolic system. Parameter identification in a diffusion operator is classically delicate \cite{richter1981,isakov2017} and in presence of noise, requires regularization \cite{engl1996,kaipio2005}, the coupling between compartments and their strong scale separation add difficulties specific to the epidemiological setting. As we make precise in Section \ref{sec:difficulties}, the difficulty is strikingly uneven across the four unknowns ($\beta,\delta,\gamma,\lambda$). 

Physics-informed neural networks (PINNs) offer an attractive route to such problems \citep{rabiei2026ai}, since the unknown state and the unknown parameters are recovered jointly from data and a governing residual \cite{raissi2019,karniadakis2021,lu2021deepxde,cuomo2022}. In epidemiology, they have been used to identify time-dependent contact rates \citep{grimm2022}, to study identifiability in integer- and fractional-order compartmental models \citep{kharazmi2021nat}, and to fit case data while inferring parameter dynamics \cite{millevoi2024}. A constraint-aware formulation for SEIR reaction-diffusion models with vital dynamics has recently been proposed \cite{zinihi2026} and serves as the matched baseline here. A standard strong-form PINN imposes the equations pointwise, which for a reaction-diffusion model requires $\lambda\Delta u_\Theta$ and hence second derivatives of the network. This is precisely the operation that makes $\lambda$ fragile: the reaction parameters multiply the field \emph{values} ($\beta SI,\ \delta E,\ \gamma I$), whereas $\lambda$ multiplies the field's \emph{curvature} $\Delta u$, so it is identified through how sharply the fields bend and not through their magnitude. A network can approximate $u$ accurately in $L^2$ while producing a biased or oscillatory $\Delta u_\Theta$, and a mean-squared field loss constrains neither the high-frequency content nor the curvature. Related PINN training pathologies like loss imbalance, spectral bias and stiff optimisation are well documented \citep{wang2021,wang2022ntk,krishnapriyan2021}.

This paper looks into variational framework for the solution. Weak and integral formulations have a long record in data-driven model discovery, precisely because they replace differentiation of noisy data by integration against smooth test functions \citep{brunton2016,schaeffer2017,reinbold2020,messenger2021}. On the neural side, the Deep Ritz method \cite{weinan2018deepritz}, variational PINNs \cite{kharazmi2019vpinn,kharazmi2021hpvpinn}, and weak adversarial networks \citep{zang2020wan,bao2020wan} exploit the same integration-by-parts technique. What has not been established is the value of the weak form for a \emph{coupled, scale-separated, noise-sensitive} inverse reaction-diffusion problem of the kind that arises in spatial epidemiology, and in particular whether it addresses the specific parameter, i.e. the diffusivity, that is the bottleneck there. 

We refer to the proposed scale-balanced variational PINN as sb-VPINN and to the scale-balanced strong-form PINN baseline as sb-PINN. Here, ``variational'' denotes the use of spatially integrated weak residuals tested against prescribed functions. In this paper, we provide:
\begin{itemize}
    \item an empirical investigation of compartment-scale imbalance reaction/diffusion-coefficient recovery, and parameter-dependent noise
    sensitivity in the inverse SEIR reaction-diffusion problem (Section \ref{sec:difficulties}), showing that diffusion remains
    the principal recovery difficulty for the scale-balanced neural estimators in the experiments considered,

    \item a scale-balanced variational physics-informed neural network (sb-VPINN) using cosine test functions and first derivatives of the network, with homogeneous Neumann conditions incorporated naturally into the weak formulation (Section \ref{sec:method}),

    \item a complementary direct weak-inversion diagnostic that reduces parameter estimation for a fixed field to a nonnegative linear least-squares problem, providing a clean-field recovery benchmark and a means of examining noise sensitivity and consistency between reconstructed fields and learned parameters, and

   \item a systematic numerical evaluation over five seeds and six noise levels in one and two spatial dimensions, demonstrating improved diffusion-coefficient recovery relative to the strong-form PINN while maintaining accurate reaction-rate estimates. The evaluation examines both clean-data accuracy and sensitivity to observation noise, showing that the benefit of the weak formulation within the PINN framework persists on a two-dimensional square domain.
\end{itemize}

\section{The inverse problem and its conditioning}
\label{sec:model}

\subsection{Model}

Let $\Omega_d=(0,L)^d\subset\R^d$, $d\in\{1,2\}$, and $t\in[0,T]$. The SEIR reaction - diffusion model with vital dynamics and a shared diffusivity $\lambda>0$ reads

\begin{equation}
\begin{aligned}
\partial_t S &= \lambda\Delta S + \Lambda - \beta SI - \mu S,\\
\partial_t E &= \lambda\Delta E + \beta pSI - (\delta+\eta+\mu)E,\\
\partial_t I &= \lambda\Delta I + \beta(1-p)SI + \delta E - (\gamma+\mu)I,\\
\partial_t R &= \lambda\Delta R + \eta E + \gamma I - \mu R,
\end{aligned}
\label{eq:seir}
\end{equation}
with homogeneous Neumann (zero-flux) boundary conditions
\begin{equation}
\partial_n S=\partial_n E=\partial_n I=\partial_n R=0 \qquad\text{on }\partial\Omega_d ,
\label{eq:neumann}
\end{equation}
where $\partial_n$ is the outward normal derivative. Here $S, E, I, R$ are the susceptible, exposed, infectious and recovered population densities, $\Lambda$ is the recruitment rate, $\mu$ the natural death rate, $\beta$ the transmission rate, $p\in[0,1]$ the fraction of newly infected individuals entering the exposed class, $\delta$ the $E \to I$ progression rate, $\eta$ the $E \to R$ transition rate and $\gamma$ the recovery rate from $I$. The zero-flux condition models an isolated spatial domain across whose boundary no individuals move.

Writing $N=S+E+I+R$ and summing \eqref{eq:seir}, every infection, progression and recovery transfer cancels, leaving the reaction-free balance
\begin{equation}
\partial_t N = \lambda\Delta N + \Lambda - \mu N .
\label{eq:total}
\end{equation}
Equation \eqref{eq:total} decouples $\lambda$ from the reaction parameters and provides an exact conservation identity that the estimator can be made to respect. 

\subsection{The inverse problem}

The constants $(\Lambda,\mu,p,\eta)$ are assumed known, and the unknown parameter vector is $\theta=(\beta,\delta,\gamma,\lambda)\in\R^4_+$. The state
$U=(S,E,I,R)$ is observed on a space-time grid $\{(t_n,x_j): 1\le n\le N_t,\ 1\le j\le N_x\}\subset[0,T]\times\overline{\Omega}_d$, and each compartment $c\in\{S,E,I,R\}$ is corrupted by additive Gaussian noise,
\begin{equation}
\Uobs_c(t_n,x_j)=U_c(t_n,x_j)+\rho\sigma_c\xi^{(c)}_{n,j}, \qquad \xi^{(c)}_{n,j}\overset{\text{i.i.d.}}{\sim}\mathcal{N}(0,1).
\label{eq:noise}
\end{equation}
Here $\rho\ge0$ is the relative noise level and $\sigma_c$ is the standard deviation of the clean values $\{U_c(t_n,x_j)\}_{n,j}$, so the noise in each
compartment is proportional to that compartment's own spread. This matters because $S$ is several orders of magnitude larger than $E$ and $I$.

\emph{Inverse problem.} Given the noisy data $\{\Uobs_c(t_n,x_j)\}$, find an estimate $\hat\theta$ of $\theta$ such that \eqref{eq:seir}-\eqref{eq:neumann} is consistent with the observations.

\emph{Error measures.} Parameter accuracy is measured componentwise by
\begin{equation}
\RelErr(\hat\theta_i)=100\frac{|\hat\theta_i-\theta_i|}{|\theta_i|}\ (\%),
\qquad i=1,\dots,4 .
\end{equation}
Every estimator used here also returns a reconstruction of the state: for the network-based methods this is $U_\Theta=(U_{\Theta,S},U_{\Theta,E},U_{\Theta,I},U_{\Theta,R})$, the network approximation with trainable weights $\Theta$ (Section \ref{sec:method}). Its accuracy is measured against the \emph{clean} field, compartment by compartment, by the discrete relative $L^2$ error on the
grid,
\begin{equation}
\frac{\|U_{\Theta,c}-U_c\|_2}{\|U_c\|_2}
=\Bigg(\frac{\sum_{n,j}\big|U_{\Theta,c}(t_n,x_j)-U_c(t_n,x_j)\big|^2}
            {\sum_{n,j}\big|U_c(t_n,x_j)\big|^2}\Bigg)^{1/2}.
\end{equation}
Throughout, the reference data are generated with $\theta=(\beta,\delta,\gamma,\lambda)=(0.4,0.3,0.2,0.05)$ (Section \ref{sec:data}).

\subsection{Three structural difficulties}
\label{sec:difficulties}

\paragraph{D1: compartment-scale heterogeneity.}
In the regime considered, most of the population remains susceptible while the exposed, infectious and removed densities are one to two orders of magnitude smaller. The clean-field root-mean-square scales are $s_S=3.5234$, $s_E=2.8906\times10^{-2}$, $s_I=8.1911\times10^{-2}$ and $s_R=1.7329\times10^{-2}$, so $E,I,R$ carry $0.82\%$, $2.32\%$ and $0.49\%$ of the susceptible scale. If each compartment carries a comparable \emph{relative} error, $U_{\Theta,c}-U_c\approx\alpha U_c$, its unweighted squared contribution scales as $\alpha^2 s_c^2$, so the susceptible compartment owns the fraction 
\begin{equation}
\frac{s_S^2}{s_S^2+s_E^2+s_I^2+s_R^2}=99.94\%
\label{eq:imbalance}
\end{equation}
of the objective: an unweighted least-squares fit is effectively optimised for $S$ alone, while ($E,I,R$) which carry the signal identifying $\beta,\delta,\gamma$ and the spatial structure identifying $\lambda$ sit in its gradient noise floor. This is an instance of the loss-imbalance pathology analysed in \cite{wang2021,wang2022ntk}. It is removable and Table \ref{tab:ablation} shows that compartment-wise normalisation (Section \ref{sec:scale}) reduces the errors in $\delta$ and $\gamma$ from $20.51\%$ and $7.97\%$ to $0.66\%$ and $0.10\%$, respectively.

\begin{table}[ht]
\centering
\caption{Effect of scale balancing on the strong-form estimator at a representative run. Normalisation resolves the reaction parameters and leaves $\lambda$ an order of magnitude worse, isolating diffusion as the residual difficulty. Field entries are relative $L^2$ errors. Relative errors are computed at full precision; recovered values are rounded to four decimals.}
\label{tab:ablation}
\small
\begin{tabular}{llcccccc}
\toprule
& & & \multicolumn{2}{c}{strong form PINN without scale balancing} & & \multicolumn{2}{c}{sb-PINN}\\
\cmidrule(lr){4-5}\cmidrule(lr){7-8}
Type & Quantity & True & Recovered / Rel.\ $L^2$ & RelErr (\%) & & Recovered / Rel.\ $L^2$ & RelErr (\%)\\
\midrule
\multirow{4}{*}{Parameter}
 & $\beta$   & 0.4000 & 0.4013 & 0.32  & & 0.3992 & 0.19 \\
 & $\delta$  & 0.3000 & 0.3615 & 20.51 & & 0.2980 & 0.66 \\
 & $\gamma$  & 0.2000 & 0.2159 & 7.97  & & 0.2002 & 0.10 \\
 & $\lambda$ & 0.0500 & 0.0262 & 47.54 & & 0.0261 & 47.75 \\
\midrule
\multirow{4}{*}{Field}
 & $S$ & - & $1.3837\times10^{-4}$ & - & & $2.3484\times10^{-3}$ & - \\
 & $E$ & - & $7.5543\times10^{-2}$ & - & & $2.7505\times10^{-2}$ & - \\
 & $I$ & - & $1.8040\times10^{-2}$ & - & & $6.2440\times10^{-3}$ & - \\
 & $R$ & - & $3.5203\times10^{-2}$ & - & & $8.2567\times10^{-3}$ & - \\
\bottomrule
\end{tabular}
\end{table}

\paragraph{D2: diffusion recovery requires derivative accuracy.}
In the strong-form residual, $\beta,\delta,\gamma$ multiply field values, whereas $\lambda$ multiplies $\Delta u_\Theta$. Writing $e=u_\Theta-u$, the error in the diffusion term is
\[
\widehat\lambda\Delta u_\Theta-\lambda\Delta u
=(\widehat\lambda-\lambda)\Delta u
+\widehat\lambda\Delta e.
\]
Thus, errors in the reconstructed Laplacian can interfere with the estimation of $\lambda$. A small field error
$\|e\|_{L^2}$ does not control $\|\Delta e\|_{L^2}$, and compartment-wise rescaling alone provides no such control.

The clean-data results show that this difficulty persists across random seeds. Over five runs, the scale-balanced strong-form PINN (sb-PINN) has a relative error of $50.35\pm9.74\%$ in $\lambda$, whereas its mean reaction-parameter errors are at most $1.10\%$ (Table \ref{tab:noise}). The sb-VPINN reduces the diffusion error to $13.72\pm4.83\%$, while keeping its mean reaction-parameter errors below $0.3\%$. This corresponds to a reduction of approximately $72.7\%$ in the mean diffusion error. These results are consistent with sensitivity to spatial-derivative accuracy, although approximation and optimisation errors may also contribute. Here, $\pm$ denotes the sample standard deviation over
five seeds.

\paragraph{D3: noise sensitivity depends on the parameter and estimator.}
In the one-dimensional experiments, observation noise affects the four parameter estimates differently. For direct weak inversion, linear fits of the seed-mean relative errors against the noise level give slopes of $0.811$, $6.199$, $1.928$, and $4.010$ percentage points of error per percentage point of noise for $\beta,\delta,\gamma,\lambda$, respectively ($R^2\geq0.944$).\footnote{For each parameter, ordinary least squares is used to fit $E=a+b r$, where $E$ is the mean relative error in percent over five seeds and $r=100\rho\in\{0,4,8,12,16,20\}$ is the noise level expressed in percent. The reported slope $b$ summarises the increase in error over the tested range.}
Thus, $\delta$ deteriorates most rapidly under noise,
despite $\lambda$ being the main recovery difficulty for
the neural estimators.

For both neural estimators, the mean reaction-rate errors remain at or below $1.5\%$ throughout the sweep, while diffusion errors remain substantially larger
(Table \ref{tab:noise}). However, these diffusion errors are already present on clean data: The strong-form mean diffusion error changes from $50.35\%$ to
$43.87\%$ between zero and $20\%$ noise, whereas the weak-form mean
error changes from $13.72\%$ to $18.55\%$. Limited deterioration under noise therefore does not by itself imply accurate recovery. Thus, the change from clean-data to affect under noise as well as the accuracy level for any data is to be analyzed.

\section{Synthetic data by a $P_1$ finite-element IMEX scheme} \label{sec:data}
Reference solutions are generated by a finite-element solver written uniformly for $d\in\{1,2\}$, so that the same assembly serves the interval and the triangulated square. Let $\mathcal{T}_h$ be a conforming mesh of $\Omega_d$ and $V_h=\{v\in C^0(\overline\Omega_d): v|_K\in P_1(K)\ \forall K\in\mathcal{T}_h\}$ with nodal basis $\{\phi_i\}_{i=1}^{N_h}$. The mass and stiffness matrices are
\begin{equation}
M_{ij}=\int_{\Omega_d}\phi_i\phi_j\mathrm{d}z, \qquad K_{ij}=\int_{\Omega_d}\nabla\phi_i\cdot\nabla\phi_j\mathrm{d}z .
\end{equation}
The Neumann condition \eqref{eq:neumann} is natural in the weak form, so no boundary degrees of freedom are removed. Writing each equation as $\partial_t U=\lambda\Delta U+F_U(S,E,I,R)$, multiplying by $v_h\in V_h$, integrating over $\Omega_d$ and integrating the diffusion term by parts gives, with $v_h=\phi_i$, the semi-discrete system 
\begin{equation}
M\frac{\mathrm{d}\mathbf U}{\mathrm{d}t}+\lambda K\mathbf U
= M\mathbf F_U(S,E,I,R),
\qquad [\mathbf F_U]_i=F_U(S_i,E_i,I_i,R_i),
\label{eq:semidiscrete}
\end{equation}
in which the boundary integral from Green's formula vanishes under \eqref{eq:neumann} and the reaction is evaluated nodally. A first-order IMEX scheme which is diffusion implicit, reaction explicit on a uniform time grid with step $\Delta t$ gives
\begin{equation}
A\mathbf U^{n+1}=M\big[\mathbf U^{n}+\Delta t\mathbf F^{n}_U\big],
\qquad A:=M+\Delta t\lambda K .
\label{eq:imex}
\end{equation}
Because $\lambda$ is shared across the four compartments, $A$ is identical for all of them and is assembled and factorised once per run. Summing the four updates cancels every internal transfer term and yields $A\mathbf N^{n+1}=M[\mathbf N^{n}+\Delta t(\Lambda\mathbf 1-\mu\mathbf N^{n})]$, the IMEX analogue of \eqref{eq:total}, which we use as a discrete conservation check.

The exposed and infectious populations are initialised as Gaussian bumps of width $\varsigma$ centred at an outbreak site $z_c\in\Omega_d$, the susceptible population uniformly, and $R(\cdot,0)=0$:
\begin{equation}
E(z,0)=E_{\mathrm{peak}}e^{-\|z-z_c\|^2/2\varsigma^2},\quad
I(z,0)=I_{\mathrm{peak}}e^{-\|z-z_c\|^2/2\varsigma^2},\quad
S(z,0)=S_0,\quad R(z,0)=0 .
\end{equation}
Unlike nonstandard finite-difference schemes, the explicit reaction treatment is not structurally positivity-preserving; an optional nodewise projection $U_i^{n+1}\leftarrow\max\{U_i^{n+1},0\}$ and an optional carrying-capacity rescaling at $N^\ast=\Lambda/\mu$ are therefore available as post-processing safeguards. Both are properties of the post-processing, not of the discretisation. The stored dataset is $\mathcal{D}_h=\{(t_n,z_i,S_i^n,E_i^n,I_i^n,R_i^n)\}$; the inverse problem is posed not against an analytical solution but against this numerical solution of the known model at prescribed parameters, with noise added by \eqref{eq:noise}.

\section{The scale-balanced variational PINN (sb-VPINN)}
\label{sec:method}

\paragraph{Notation.} The observed data of Section \ref{sec:data} are stored on
the space-time grid
\begin{equation*}
\mathcal{G}=\{(t_n,x_j):\ n=1,\dots,N_t,\ j=1,\dots,N_x\},
\end{equation*}
where $N_t$ is the number of stored time levels and $N_x$ the number of spatial nodes, and $\mathcal{G}$ contains $N_tN_x$ points. The estimator uses four sets of points: $N_d$ data points (data loss), the $N_x$ nodes of the initial slice (initial-condition loss), $N_\tau$ time levels (weak PDE loss) and $N_r$ constraint points (non-negativity and population losses). The weak residual is tested against $N_\varphi$ cosine functions. The subscript $c\in\{S,E,I,R\}$ always denotes a compartment.

\subsection{Neural representation and admissibility}

The state is represented by a fully connected $\tanh$ network $U_\Theta:[0,T]\times\Omega\to\R^4$ of depth $6$ and width $64$, with Xavier-normal weights and zero biases, whose inputs are mapped affinely to $[-1,1]$, $\hat t=2(t-t_{\min})/(t_{\max}-t_{\min})-1$ and analogously for $\hat x$, so that they remain in the active range of $\tanh$. The unknown parameters are optimised as unconstrained raw variables $(\tilde\beta,\tilde\delta,\tilde\gamma,\tilde\lambda)$ mapped to admissible values by $\beta=\mathrm{softplus}(\tilde\beta)$ and so on, which imposes positivity structurally rather than through a penalty. The raw variables are chosen so that the transformed parameters start at prescribed values distinct from the ground truth; for example, (for instance $\lambda_{\text{init}}=0.02$  against a true $0.05$), so that an accurate final estimate demonstrates learning rather than a fortuitous initialisation.

\subsection{Scale balancing}
\label{sec:scale}

The compartments have different magnitudes, so an unweighted squared-error loss can be dominated by the largest compartment. We account for these differences both in the network outputs and in the loss. For each compartment $c\in\{S,E,I,R\}$, we compute
\begin{equation}
s_c=
\left(
\frac{1}{N_tN_x}
\sum_{n=1}^{N_t}\sum_{j=1}^{N_x}
\Uobs_c(t_n,x_j)^2
\right)^{1/2}
+\varepsilon,
\qquad \varepsilon=10^{-12}.
\label{eq:scale}
\end{equation}
These scales are computed once from the observations on the full grid $\mathcal G$, before sampling points for the data loss, and remain fixed during training. The clean synthetic fields are not used. The small constant $\varepsilon$ prevents division by zero. The scale is used in two places, which act on different objects.

\emph{Scaling the outputs.}
Writing $\widetilde U_\Theta$ for the raw network output, we recover the physical field as
\begin{equation}
U_{\Theta,c}(t,x)=s_c\widetilde U_{\Theta,c}(t,x).
\label{eq:outscale}
\end{equation}
The target for each network output therefore has approximately unit root-mean-square magnitude. The model equations and constraints are evaluated using the physical field $U_\Theta$.

\emph{Scaling the losses.}
Each compartment residual is divided by the same scale before it is squared. With the averaging conventions used in \eqref{eq:total-loss},
\begin{equation}
\mathcal L_\ell
=
\begin{cases}
\displaystyle
\frac{1}{4N_\ell}
\sum_{c\in\{S,E,I,R\}}
\left\|\frac{r_c^{(\ell)}}{s_c}\right\|_2^2,
& \ell=\text{data or initial condition},\\[2ex]
\displaystyle
\frac{1}{N_\ell}
\sum_{c\in\{S,E,I,R\}}
\left\|\frac{r_c^{(\ell)}}{s_c}\right\|_2^2,
& \ell=\text{PDE or non-negativity}.
\end{cases}
\label{eq:lossscale}
\end{equation}
Here $N_\ell$ is the number of residual entries per compartment:
$N_d$ for the data, $N_x$ for the initial condition,
$N_\tau N_\varphi$ for the weak PDE residual, and $N_r$ for
non-negativity. 

\emph{How the two operations work together.}
Output scaling gives the network comparable target magnitudes; loss weighting measures each compartment's error relative to its own field scale. For a single data residual, their combination gives
\begin{equation}
\begin{aligned}
\frac{(U_{\Theta,c}-\Uobs_c)^2}{s_c^2}
&=
\left(\widetilde U_{\Theta,c}-\frac{\Uobs_c}{s_c}\right)^2,\\
\frac{\partial}{\partial\widetilde U_{\Theta,c}}
\frac{(U_{\Theta,c}-\Uobs_c)^2}{s_c^2}
&=
2\left(\widetilde U_{\Theta,c}-\frac{\Uobs_c}{s_c}\right).
\end{aligned}
\label{eq:gradfactor}
\end{equation}
Thus the gradient at the output depends only on the scale-normalised error, with no additional factor of $s_c$. The same cancellation applies to the initial-condition and non-negativity penalties. It does not imply equal gradients for all network parameters, since these also depend on the network derivatives and the coupled PDE.

\emph{Effect of observation noise.}
We compare the scale computed from noisy observations with the RMS of the clean field,
\[
\bar s_c^{2}
=
\frac{1}{N_tN_x}\sum_{n,j}U_c(t_n,x_j)^2.
\]
The clean field is used only for this analysis, not to compute the training scales.

By \eqref{eq:scale} and the additive noise model \eqref{eq:noise},
\[
(s_c-\varepsilon)^2=\frac{1}{N_tN_x}\sum_{n,j}\left(U_c(t_n,x_j)+\rho\sigma_c\xi_{n,j}^{(c)}\right)^2.
\]
Taking expectation over the observation noise, the cross terms vanish because $\mathbb E[\xi_{n,j}^{(c)}]=0$, while
$\mathbb E[(\xi_{n,j}^{(c)})^2]=1$. Hence
\begin{equation}
\mathbb E \left[(s_c-\varepsilon)^2\right]=\bar s_c^{2}+\rho^2\sigma_c^2.
\label{eq:scale-noise}
\end{equation}
Here $\varepsilon$ has been retained exactly: $s_c-\varepsilon$ is the RMS of the noisy observations.

When the clean-field standard deviation $\sigma_c$ is computed with divisor $N_tN_x$, it satisfies $\sigma_c^2\leq\bar s_c^{2}$. Therefore, for $\bar s_c>0$,
\begin{equation}
0\leq\frac{\mathbb E[(s_c-\varepsilon)^2]-\bar s_c^{2}}{\bar s_c^{2}}=\rho^2\frac{\sigma_c^2}{\bar s_c^{2}}\leq \rho^2.
\label{eq:scale-noise-bound}
\end{equation}
Thus, at $\rho=0.20$, the expected squared RMS increases by at most
$4\%$. This is an expectation over noise realisations, not a
deterministic bound for an individual dataset. At $\rho=0$,
the scale is exactly $s_c=\bar s_c+\varepsilon$.

\subsection{The weak residual}

For $m=0,1,\dots,N_\varphi-1$, where $N_\varphi$ is the number of
test functions, let
\begin{equation}
\varphi_m(x)=\cos\Big(\frac{m\pi x}{L}\Big),
\qquad
\varphi_m'(x)=-\frac{m\pi}{L}\sin\Big(\frac{m\pi x}{L}\Big)
\end{equation}
be cosine test functions compatible with the homogeneous Neumann condition.

\begin{proposition}
\label{prop:ibp}
Let $u\in H^2(\Omega)$ satisfy $\partial_n u=0$ on $\partial\Omega=\{0,L\}$.
Then for every $m$,
\begin{equation}
-\lambda\int_\Omega u_{xx}\varphi_m\mathrm{d}x
= -\lambda\big[u_x\varphi_m\big]_{\partial\Omega}
  +\lambda\int_\Omega u_x\varphi_m'\mathrm{d}x
= \lambda\int_\Omega u_x\varphi_m'\mathrm{d}x .
\label{eq:ibp}
\end{equation}
\end{proposition}

\noindent
The proof is a single integration by parts; the boundary contribution vanishes because
$u_x=0$ at both endpoints. Three consequences follow. First, the diffusion coefficient is
now identified through a first spatial derivative and a spatial average, not through
pointwise curvature: the network is never asked for $U_{\Theta,xx}$, which is the operation
behind D2. Second, the network itself is not required to satisfy
$\partial_nU_\Theta=0$ pointwise. Since $\varphi_m(0)=1$ and $\varphi_m(L)=(-1)^m$ do not
vanish, the residuals \eqref{eq:weakres} use the right-hand side of \eqref{eq:ibp}, without the boundary term. A smooth field that satisfies them for every test function in $H^1(\Omega)$
satisfies the equations in $\Omega$ and, on integrating by parts back, $\partial_n u=0$ on
$\partial\Omega$. The Neumann condition is therefore natural: it is imposed weakly through the
test space, and with $N_\varphi$ modes only in this projected sense, but no boundary penalty
and no boundary weight are needed. Third, for $m=0$, $\varphi_0\equiv1$ and
$\varphi_0'\equiv0$, so the diffusion term drops out and each residual reduces to the
spatially integrated balance of its compartment; their sum is the integrated form of
\eqref{eq:total}. This mode constrains the reaction terms and the total mass, but carries no
information about $\lambda$.

With $\langle f\rangle_m=\int_\Omega f\varphi_m\mathrm{d}x$ and
$\langle f_x,\varphi_m'\rangle=\int_\Omega f_x\varphi_m'\mathrm{d}x$, the
weak residuals at a time level $t$ are
\begin{equation}
\begin{aligned}
r_S(t,m)&=\langle S_{\Theta,t}\rangle_m-\Lambda\langle1\rangle_m
  +\mu\langle S_\Theta\rangle_m+\beta\langle S_\Theta I_\Theta\rangle_m
  +\lambda\langle S_{\Theta,x},\varphi_m'\rangle,\\
r_E(t,m)&=\langle E_{\Theta,t}\rangle_m+(\delta+\eta+\mu)\langle E_\Theta\rangle_m
  -\beta p\langle S_\Theta I_\Theta\rangle_m+\lambda\langle E_{\Theta,x},\varphi_m'\rangle,\\
r_I(t,m)&=\langle I_{\Theta,t}\rangle_m+(\gamma+\mu)\langle I_\Theta\rangle_m
  -\beta(1-p)\langle S_\Theta I_\Theta\rangle_m-\delta\langle E_\Theta\rangle_m
  +\lambda\langle I_{\Theta,x},\varphi_m'\rangle,\\
r_R(t,m)&=\langle R_{\Theta,t}\rangle_m-\eta\langle E_\Theta\rangle_m
  -\gamma\langle I_\Theta\rangle_m+\mu\langle R_\Theta\rangle_m
  +\lambda\langle R_{\Theta,x},\varphi_m'\rangle .
\end{aligned}
\label{eq:weakres}
\end{equation}
Spatial integrals are approximated by the trapezoidal rule on the
$N_x$ spatial nodes; the first derivatives
$U_{\Theta,t},U_{\Theta,x}$ are obtained by automatic differentiation.
The residuals are evaluated at $N_\tau$ time levels
$t_\tau^{(1)},\dots,t_\tau^{(N_\tau)}$ selected from the stored levels $t_n$. %so that each residual entry uses a whole spatial slice rather than a single point. 
The weak PDE loss carries the weighting of \eqref{eq:lossscale},
\begin{equation}
\mathcal{L}^{\mathrm{weak}}_{\mathrm{PDE}}
=\sum_{c\in\{S,E,I,R\}}\frac{1}{s_c^2}
\frac{1}{N_\tau N_\varphi}
\sum_{q=1}^{N_\tau}\sum_{m=0}^{N_\varphi-1}\big|r_c(t_\tau^{(q)},m)\big|^2 .
\label{eq:weakloss}
\end{equation}

\subsection{Objective and staged training}

The total loss is
\begin{equation}
\mathcal{L}_{\mathrm{total}}
=w_{\mathrm{data}}\mathcal{L}_{\mathrm{data}}
+w_{\mathrm{IC}}\mathcal{L}_{\mathrm{IC}}
+w_{\mathrm{phys}}\mathcal{L}^{\mathrm{weak}}_{\mathrm{PDE}}
+w_{\mathrm{nonneg}}\mathcal{L}_{\mathrm{nonneg}}
+w_{\mathrm{pop}}\mathcal{L}_{\mathrm{pop}} .
\label{eq:total-loss}
\end{equation}
The data loss is evaluated on $N_d$ points
$(t_d^{(q)},x_d^{(q)})$, $q=1,\dots,N_d$, sampled from $\mathcal{G}$, and the
initial-condition loss on the $N_x$ nodes of the observed slice at $t=0$:
\begin{equation}
\mathcal{L}_{\mathrm{data}}=\frac{1}{4N_d}\sum_{q=1}^{N_d}\sum_{c}
\frac{|U_{\Theta,c}(t_d^{(q)},x_d^{(q)})-\Uobs_c(t_d^{(q)},x_d^{(q)})|^2}{s_c^2},
\qquad
\mathcal{L}_{\mathrm{IC}}=\frac{1}{4N_x}\sum_{j=1}^{N_x}\sum_c
\frac{|U_{\Theta,c}(0,x_j)-\Uobs_c(0,x_j)|^2}{s_c^2}.
\end{equation}
The two constraint losses are evaluated on $N_r$ points
$(t_r^{(q)},x_r^{(q)})$, $q=1,\dots,N_r$, sampled at random in
$[0,T]\times\Omega$. With $N_\Theta=S_\Theta+E_\Theta+I_\Theta+R_\Theta$ the
total population predicted by the network and $N^\ast=\Lambda/\mu$ the carrying
capacity,
\begin{equation}
\mathcal{L}_{\mathrm{nonneg}}=\frac{1}{N_r}\sum_{q=1}^{N_r}\sum_c
\frac{[\max(0,-U_{\Theta,c}(t_r^{(q)},x_r^{(q)}))]^2}{s_c^2},
\qquad
\mathcal{L}_{\mathrm{pop}}=\frac{1}{N_r}\sum_{q=1}^{N_r}
\frac{[\max(0,N_\Theta(t_r^{(q)},x_r^{(q)})-N^\ast)]^2}{(N^\ast)^2}.
\end{equation}
All inter-term weights $w_\bullet$ are set to unity; the only non-uniform weighting in the method is the division by $s_c^2$ within each term. The
initial condition is anchored to the \emph{observed} initial data rather than
to the clean prescribed state, which would not be available with
real data. There is no boundary loss: Proposition \ref{prop:ibp} makes the
Neumann condition natural.

Because $\theta$ enters \eqref{eq:total-loss} only through
$\mathcal{L}^{\mathrm{weak}}_{\mathrm{PDE}}$, in any phase in which that term is
inactive the parameters receive no gradient and remain at their initial values.
This licenses the staged schedule of Algorithm \ref{alg:train}: activating the
physics from the start would let $\theta$ chase a field that is still
uninformative.

\begin{algorithm}[ht]
\caption{sb-VPINN training for parameter recovery}
\label{alg:train}
\begin{algorithmic}[1]
\State \textbf{Input:} observations $\Uobs$, known constants
$(\Lambda,\mu,p,\eta)$.
\State Estimate the scales $s_c$ from \eqref{eq:scale}; initialise network
weights (Xavier) and raw parameters $\tilde\theta$ at admissible, off-truth
values.
\State \textbf{Stage A (warm-up):} minimise
$w_{\mathrm{data}}\mathcal{L}_{\mathrm{data}}+w_{\mathrm{IC}}\mathcal{L}_{\mathrm{IC}}
+\mathcal{L}_{\mathrm{constraints}}$ by Adam with the weak PDE residual
switched off. The parameters receive no gradient and stay at their initial
values.
\State \textbf{Stage B (joint):} with a fresh Adam optimiser (cosine-annealed
step size, gradient clipping) and the weak PDE residual active, co-train the
network weights $\Theta$ and the parameters $\theta$ on
$\mathcal{L}_{\mathrm{total}}$.
\State \textbf{Stage C (polish):} apply a deterministic L-BFGS refinement
(strong-Wolfe line search) to the full objective.
\State \textbf{Output:} $\theta=\mathrm{transform}(\tilde\theta)$ and the
reconstructed field $U_\Theta$.
\end{algorithmic}
\end{algorithm}

\begin{figure}[htbp]
\centering
\includegraphics[width=0.92\linewidth]{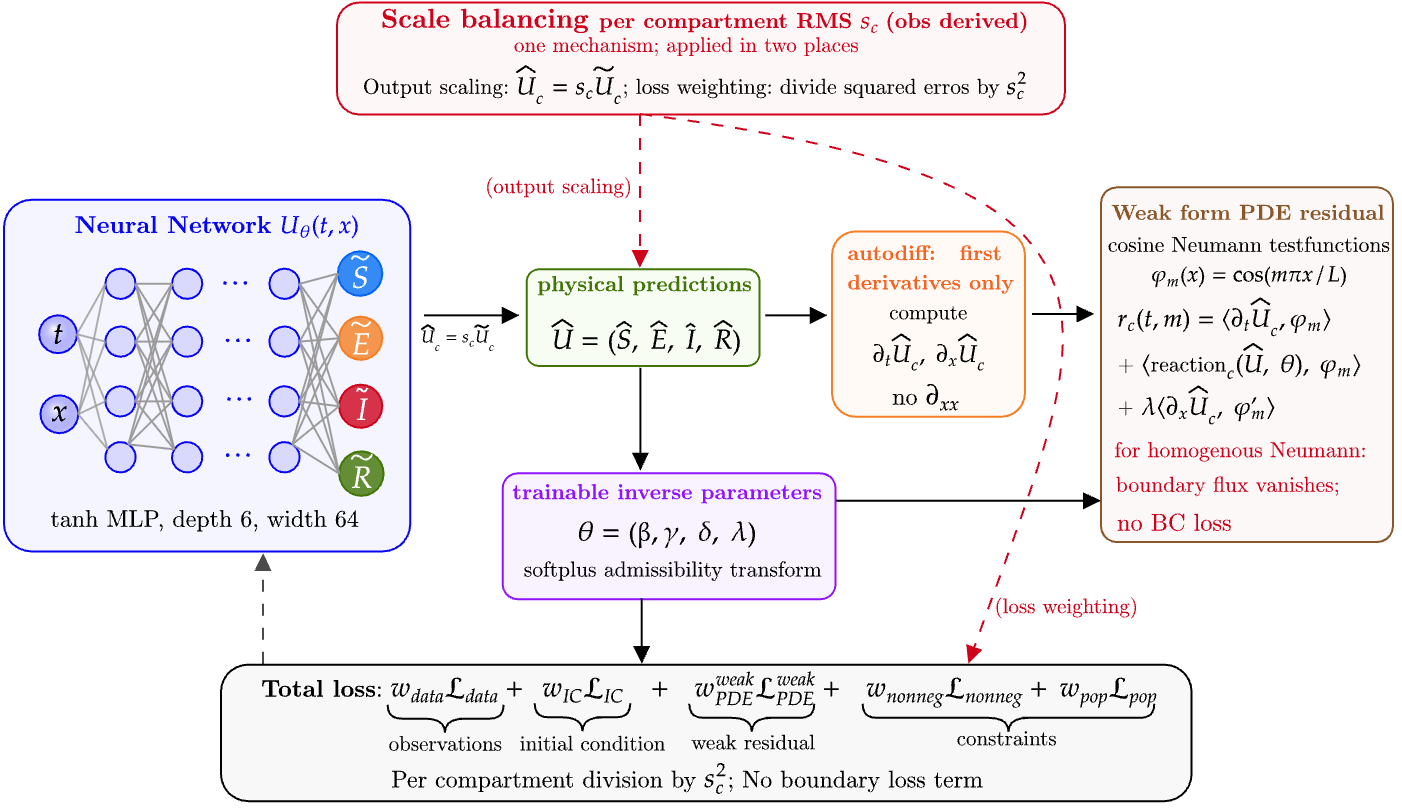}
\caption{Architecture of sb-VPINN. The per-compartment root-mean-square scale $s_c$, computed from the observed field is applied in two places, as an output non-dimensionalisation\eqref{eq:outscale} and as a loss-side divisor \eqref{eq:lossscale}. The physics residual is imposed in weak form against cosine Neumann test functions. By Proposition \ref{prop:ibp} only first derivatives of the network are formed and no boundary loss is required. The unknown parameters enter the objective only through the weak residual. }
\label{fig:schematic}
\end{figure}

\subsection{The direct weak inversion diagnostic}
\label{sec:direct}

The weak residuals \eqref{eq:weakres} also allow parameter recovery
without a neural network when the field $U=(S,E,I,R)$ is supplied.
Replacing $U_\Theta$ by $U$, each residual is affine in
$\theta=(\beta,\delta,\gamma,\lambda)^{\mathsf T}$.
For example, setting the susceptible-compartment residual to zero gives
\begin{equation}
\beta\langle SI\rangle_m
+\lambda\langle S_x,\varphi_m'\rangle
=
\Lambda\langle1\rangle_m
-\mu\langle S\rangle_m-\langle S_t\rangle_m.
\label{eq:direct-S}
\end{equation}
All quantities other than $\beta$ and $\lambda$ are determined by
the supplied field and the known coefficients. The other three
equations have the same linear dependence on the unknown parameters.

We use the same $N_\varphi$ test functions as in the weak PDE loss,
evaluated at $N_\tau^{\mathrm{dir}}$ approximately evenly spaced
stored time indices. The superscript distinguishes this number
from the $N_\tau$ time levels used in training.
Stacking the residuals over compartments, times and modes gives
\[
r(U;\theta)
=
A_{\mathrm{weak}}(U)\theta-b_{\mathrm{weak}}(U).
\]
Writing $A_c(U)$ and $b_c(U)$ for the rows associated with compartment
$c$, we formulate the direct estimate as
\begin{equation}
\hat\theta_{\mathrm{direct}}
\in
\operatorname*{argmin}_{\theta\in[0,\infty)^4}
\sum_{c\in\{S,E,I,R\}}
\frac{\|A_c(U)\theta-b_c(U)\|_2^2}{s_c^2}.
\label{eq:direct}
\end{equation}
The same observation-derived scales $s_c$ are used for both the
clean-field and observed-field calculations. Thus the direct
system uses the same compartment weighting as the neural weak
PDE loss.

The spatial integrals are approximated by the trapezoidal rule.
Unlike training, the direct calculation obtains $U_t$ and $U_x$
by finite differences on the supplied grid arrays, before selecting
the times used in the system. It therefore uses grid information
beyond the $N_d$ sampled data-loss points.
The resulting weighted problem is passed to a non-negative
least-squares solver.\footnote{If this solver fails, the supplied
implementation falls back to unconstrained least squares followed
by clipping negative components to zero. That fallback does not,
in general, solve \eqref{eq:direct}.}

We report the diagnostic on the \emph{clean synthetic field} as a
noise-free numerical reference, and on the \emph{noisy observed field}
to assess sensitivity to measurement noise. The clean-field result
includes discretisation and numerical differentiation errors. For the observed field, noise perturbs both $A_{\mathrm{weak}}(U)$ and $b_{\mathrm{weak}}(U)$.
Spatial averaging avoids pointwise second derivatives but does not
eliminate this sensitivity. The direct estimates are reported separately from the trained parameters and are not used to update the network or its coefficients.

% =============================================================================
\section{Numerical results}
% =============================================================================

\subsection{Protocols}

Experiments use the finite-element data of Section \ref{sec:data} on $\Omega=(0,L)$, $t\in[0,T]$, with true parameters
$\beta=0.4$, $\delta=0.3$, $\gamma=0.2$, $\lambda=0.05$ and known constants $(\Lambda,\mu,p,\eta)$. Both neural estimators use the same network architecture, observation-derived compartment scaling and optimisation schedule.
The strong form evaluates the PDE residual at randomly sampled
space-time points and includes an explicit Neumann boundary penalty.
The weak form evaluates spatially integrated residuals at selected
time levels using the full spatial grid, with the Neumann condition
incorporated naturally.
%The comparison against the strong-form estimator is \emph{matched}: architecture, data, collocation, compartment scales, optimiser and training budget are identical, 
In other words, the two models differ through
\begin{equation}
-\lambda\int_\Omega\Delta U_\Theta\varphi_m\mathrm{d}x
=\lambda\int_\Omega\nabla U_\Theta\cdot\nabla\varphi_m\mathrm{d}x ,
\label{eq:onlydiff}
\end{equation}
that is, weak first-derivative spatial averages versus pointwise curvature $\lambda\Delta U_\Theta$. Every configuration is run over five seeds ($0$-$4$) at six noise levels $\rho\in\{0,4,8,12,16,20\}\%$; tables report mean $\pm$ sample standard deviation and the per-seed record is given in Appendix \ref{app:perseed}.

\begin{table}[htbp]
\centering
\caption{Experimental configuration for the one-dimensional evaluation.}
\label{tab:experimental_configuration}
\begin{tabular}{ll}
\toprule
Quantity & Value \\
\midrule
Spatial domain / time interval & $(0,1)$ / $[0,5]$ \\
Spatial nodes / time snapshots & 101 / 2001\\
True $\beta,\delta,\gamma,\lambda$ & $0.4,0.3,0.2,0.05$ \\
Known constants $\Lambda,\mu,p,\eta$ & $1.0,0.01,0.3,0.1$ \\
Network depth / width & $6$ / $64$ \\
%Collocation points & $6000$ \\
Strong-form PDE collocation points & $6000$ \\
Weak PDE time levels / spatial nodes per level & $80$ / $101$ \\
Weak test functions & $N_\varphi=8$ \\
Direct-inversion time levels / spatial nodes per level & $200$ / $101$ \\
%Weak test modes & $N_\phi=8$ \\
Warm-up / Adam / L-BFGS budget & $1000$ / $7000$ / $300$ \\
Seeds & 5 \\
Noise levels $\rho$ & $0,4,8,12,16,20\%$ \\
\bottomrule
\end{tabular}
\end{table}

Three questions organise the evaluation of sb-VPINN. \\
\textbf{Q1}: does the weak form improve clean-data recovery of $\lambda$ over the matched scale-balanced strong form?\\ 
\textbf{Q2}: does it preserve the accuracy of $\beta,\delta,\gamma$? \\
\textbf{Q3}: how does recovery behave under noise? \\
%\textbf{Q4}: is the trained field consistent with the co-trained coefficients?

\subsection{Clean data (Q1 and Q2)}

\begin{figure}[ht]
\centering
\includegraphics[scale=0.8]{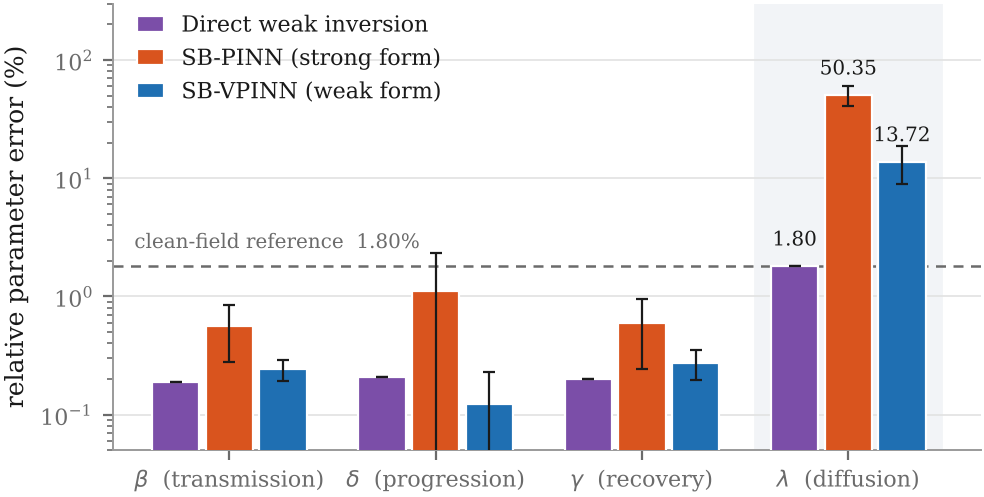}
\caption{Clean-data parameter recovery in one dimension (relative error, logarithmic scale; mean $\pm$ sample standard deviation over seeds $0$-$4$). The mean reaction-parameter errors stay below $1.2\%$ for all three estimators, and only the shaded diffusion group separates them, with mean errors in $\lambda$ of $1.80\%$ (direct weak inversion), $50.35\%$ (sb-PINN) and $13.72\%$ (sb-VPINN). The dashed line is the clean-field reference of $1.80\%$ obtained by direct weak inversion of the clean finite-element field (Section \ref{sec:direct}).}
\label{fig:clean}
\end{figure}

Clean data isolates learning bias from noise. With noise-free observations, any error is attributable to the neural approximation, the optimiser or the conditioning of the problem. Table \ref{tab:clean} compares three estimators at a representative run, and Figure \ref{fig:clean} shows the seed-aggregated picture. The scale-balanced strong-form estimator has a relative error of $47.75\%$ in $\lambda$, compared with $10.21\%$ for the weak form, a reduction by a factor of approximately $4.68$. The reaction-parameter errors are at most $0.66\%$ for the strong form and $0.30\%$ for the weak form, i.e., the weak form does not trade reaction-parameter accuracy for it. Direct weak inversion of the clean synthetic field gives a relative error of $1.80\%$ in $\lambda$, providing a numerical benchmark for recovery when the field is supplied rather than learned. Across five seeds, the diffusion errors are $50.35\pm9.74\%$ for
the scale-balanced strong-form PINN (sb-PINN) and $13.72\pm4.83\%$ for the
scale-balanced weak-form PINN (sb-VPINN),  confirming the improvement beyond the representative run.

\begin{table}[ht]
\centering
\caption{
Clean-data parameter recovery at a representative run (relative error, \%): direct inversion on the exact field, scale-balanced strong-form estimator, and the scale-balanced weak-form estimator.
}
\label{tab:clean}
\small
\begin{tabular}{lcccccccc}
\toprule
& & \multicolumn{2}{c}{Direct (clean FEM field)} & \multicolumn{2}{c}{sb-PINN}
& \multicolumn{2}{c}{sb-VPINN}\\
\cmidrule(lr){3-4}\cmidrule(lr){5-6}\cmidrule(lr){7-8}\cmidrule(lr){9-9}
Param. & True & Estimate & RelErr & Estimate & RelErr & Estimate & RelErr\\
\midrule
$\beta$   & 0.4000 & 0.3993 & 0.19 & 0.3992 & 0.19 & 0.3990 & 0.24 \\
$\delta$  & 0.3000 & 0.2994 & 0.21 & 0.2980 & 0.66 & 0.2992 & 0.28 \\
$\gamma$  & 0.2000 & 0.1996 & 0.20 & 0.2002 & 0.10 & 0.1994 & 0.30 \\
$\lambda$ & 0.0500 & 0.0491 & \textbf{1.80} & 0.0261 & \textbf{47.75} & 0.0449 & \textbf{10.21} \\
\bottomrule
\end{tabular}
\end{table}

\subsection{Field reconstruction}

An estimator may recover parameters yet reconstruct the state poorly, or fit the state well and mis-estimate $\lambda$, which depends on spatial-derivative information rather than on field values alone. Table \ref{tab:fields} reports the relative $L^2$ error per compartment for the weak-form estimator. All four compartments reach order $10^{-3}$, and the weak form is more accurate than the strong form in every compartment by factors of $1.68$, $4.76$, $3.29$ and $5.23$ for $S,E,I,R$ against the right-hand block of Table \ref{tab:ablation}. Crucially, the small compartments $E,I,R$, where the diffusion signal resides, are recovered as faithfully as $S$ despite being one to two orders of magnitude
smaller: the direct payoff of scale balancing. 
\emph{A sub-$1\%$ field accuracy alongside a $10\%$ error in $\lambda$ shows that the difficulty in the diffusion coefficient is not poor field reconstruction but the curvature sensitivity that even a highly accurate field cannot resolve.}

\begin{table}[ht]
\centering
\caption{Field reconstruction errors for the sb-VPINN estimator in a representative run. Relative $L^2$ errors range from $1.40\times10^{-3}$ to $5.78\times10^{-3}$.}
\label{tab:fields}
\small
\begin{tabular}{lcccc}
\toprule
Compartment & Rel.\ $L^2$ & MAE & RMSE & Max\\
\midrule
$S$ & $1.4005\times10^{-3}$ & $3.7680\times10^{-3}$ & $4.9346\times10^{-3}$ & $1.8192\times10^{-2}$\\
$E$ & $5.7813\times10^{-3}$ & $9.1000\times10^{-5}$ & $1.6711\times10^{-4}$ & $2.3329\times10^{-3}$\\
$I$ & $1.8978\times10^{-3}$ & $1.1220\times10^{-4}$ & $1.5545\times10^{-4}$ & $1.6083\times10^{-3}$\\
$R$ & $1.5801\times10^{-3}$ & $2.0232\times10^{-5}$ & $2.7381\times10^{-5}$ & $2.8055\times10^{-4}$\\
\bottomrule
\end{tabular}
\end{table}

\subsection{Behaviour under noise (Q3)}

\begin{figure}[ht]
\centering
\includegraphics[width=0.7\linewidth]{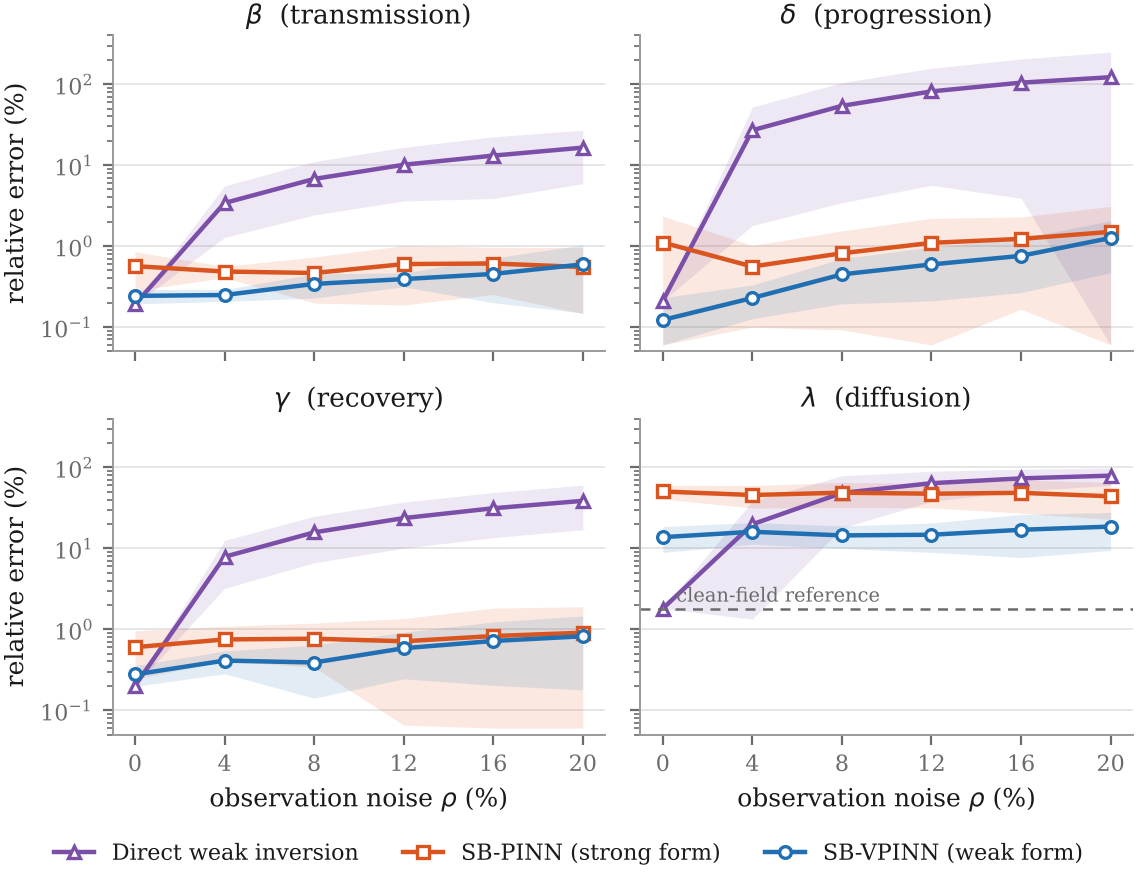}
\caption{Relative parameter error versus observation noise in one dimension (mean over seeds $0$-$4$; shaded bands show one sample standard deviation, clipped below at $0.06\%$ for display on the logarithmic axis; all panels share the same scale). The mean reaction-parameter errors of both neural estimators stay at or below $1.5\%$, whereas the direct weak inversion degrades with noise; the $\lambda$ panel separates the three estimators. The dashed line is the clean-field reference ($1.80\%$).}
\label{fig:noise}
\end{figure}

Table \ref{tab:noise} and Figure \ref{fig:noise} compare the three estimators over five seeds and six noise levels. For the weak form, the mean errors in $\beta,\delta,\gamma$ remain below $1.3\%$ throughout. Its mean error in $\lambda$ increases from $13.72\%$ on clean data to $18.55\%$ at $20\%$ noise. The strong form has a much larger
diffusion error even on clean data, with mean errors ranging from $43.87\%$ to $50.35\%$ across the noise levels. The weak form gives a smaller diffusion error in $28$ of the $30$ runs and a lower mean error at every noise level. Direct weak inversion is accurate on clean data but deteriorates substantially as noise increases: at $20\%$ noise, its mean errors reach $79.02\%$ in $\lambda$ and $121.87\%$ in $\delta$.

\begin{table}[ht]
\centering
\caption{Noise robustness: relative parameter errors (\%), mean $\pm$ sample standard deviation over seeds $0$-$4$. The per-seed record is given in Table \ref{tab:perseed}.}
\label{tab:noise}
\footnotesize
\setlength{\tabcolsep}{4.5pt}
\resizebox{\linewidth}{!}{%
\begin{tabular}{llcccccc}
\toprule
Method & Param. & Clean & $\rho=4\%$ & $\rho=8\%$ & $\rho=12\%$ & $\rho=16\%$ & $\rho=20\%$\\
\midrule
\multirow{4}{*}{Direct weak inversion}
 & $\beta$   & $0.19\pm0.00$ & $3.40\pm2.12$   & $6.74\pm4.32$   & $10.09\pm6.49$  & $13.09\pm9.23$    & $16.42\pm10.52$\\
 & $\delta$  & $0.21\pm0.00$ & $26.96\pm25.17$ & $53.94\pm50.51$ & $81.35\pm75.73$ & $104.34\pm100.46$ & $121.87\pm126.49$\\
 & $\gamma$  & $0.20\pm0.00$ & $7.86\pm4.66$   & $15.74\pm9.11$  & $23.60\pm13.58$ & $31.23\pm17.70$   & $38.58\pm21.58$\\
 & $\lambda$ & $1.80\pm0.00$ & $19.82\pm18.49$ & $48.28\pm30.42$ & $63.76\pm25.75$ & $73.11\pm21.77$   & $79.02\pm19.25$\\
\midrule
\multirow{4}{*}{sb-PINN}
 & $\beta$   & $0.56\pm0.28$  & $0.48\pm0.08$   & $0.46\pm0.27$   & $0.60\pm0.41$   & $0.61\pm0.36$   & $0.55\pm0.40$\\
 & $\delta$  & $1.10\pm1.24$  & $0.56\pm0.46$   & $0.81\pm0.72$   & $1.09\pm1.11$   & $1.23\pm1.06$   & $1.50\pm1.59$\\
 & $\gamma$  & $0.60\pm0.36$  & $0.75\pm0.34$   & $0.76\pm0.43$   & $0.71\pm0.64$   & $0.82\pm1.00$   & $0.91\pm0.99$\\
 & $\lambda$ & $50.35\pm9.74$ & $45.55\pm14.04$ & $48.66\pm16.61$ & $47.25\pm15.71$ & $48.46\pm21.21$ & $43.87\pm21.17$\\
\midrule
\multirow{4}{*}{\textbf{sb-VPINN}}
 & $\beta$   & $0.24\pm0.05$ & $0.25\pm0.04$ & $0.34\pm0.11$ & $0.39\pm0.08$ & $0.45\pm0.25$ & $0.59\pm0.45$\\
 & $\delta$  & $0.12\pm0.11$ & $0.23\pm0.10$ & $0.45\pm0.25$ & $0.59\pm0.38$ & $0.76\pm0.49$ & $1.26\pm0.79$\\
 & $\gamma$  & $0.28\pm0.08$ & $0.41\pm0.13$ & $0.39\pm0.25$ & $0.58\pm0.34$ & $0.72\pm0.51$ & $0.82\pm0.64$\\
 & $\lambda$ & $\mathbf{13.72\pm4.83}$ & $\mathbf{16.00\pm4.78}$ & $\mathbf{14.44\pm4.48}$
             & $\mathbf{14.70\pm5.79}$ & $\mathbf{16.94\pm9.25}$ & $\mathbf{18.55\pm9.14}$\\
\bottomrule
\end{tabular}}
\end{table}

\begin{figure}[htbp]
\centering
\includegraphics[width=1\linewidth]{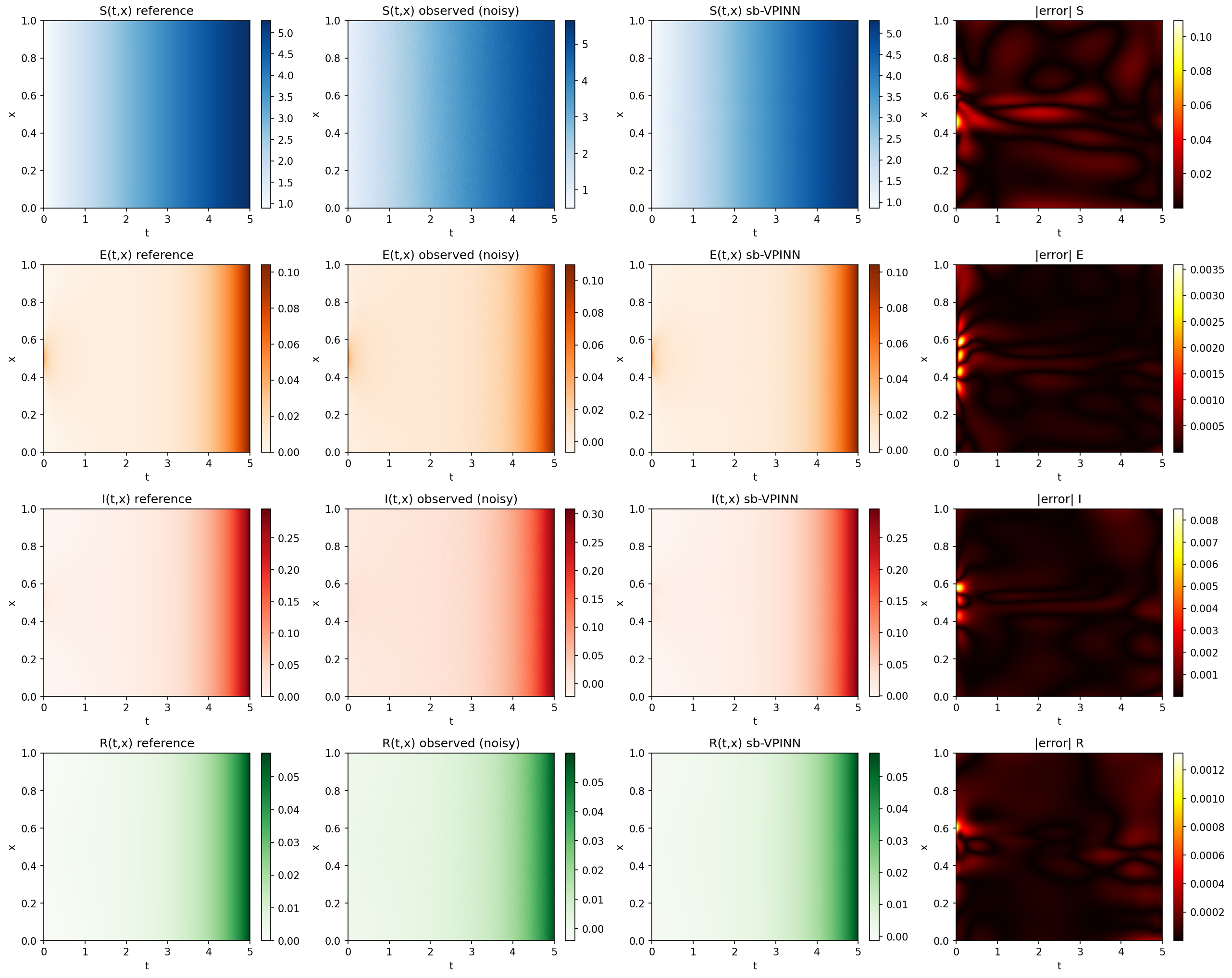}
\caption{Space-time reconstruction for the sb-VPINN estimator on observed data with $8\%$ noise. Each row shows, for one compartment, the reference (clean) field, the observed (noisy) field, the learned field and the absolute error; note the different error scales. }
\label{fig:spacetime}
\end{figure}

\begin{figure}[htbp]
\centering
\includegraphics[width=0.92\linewidth]{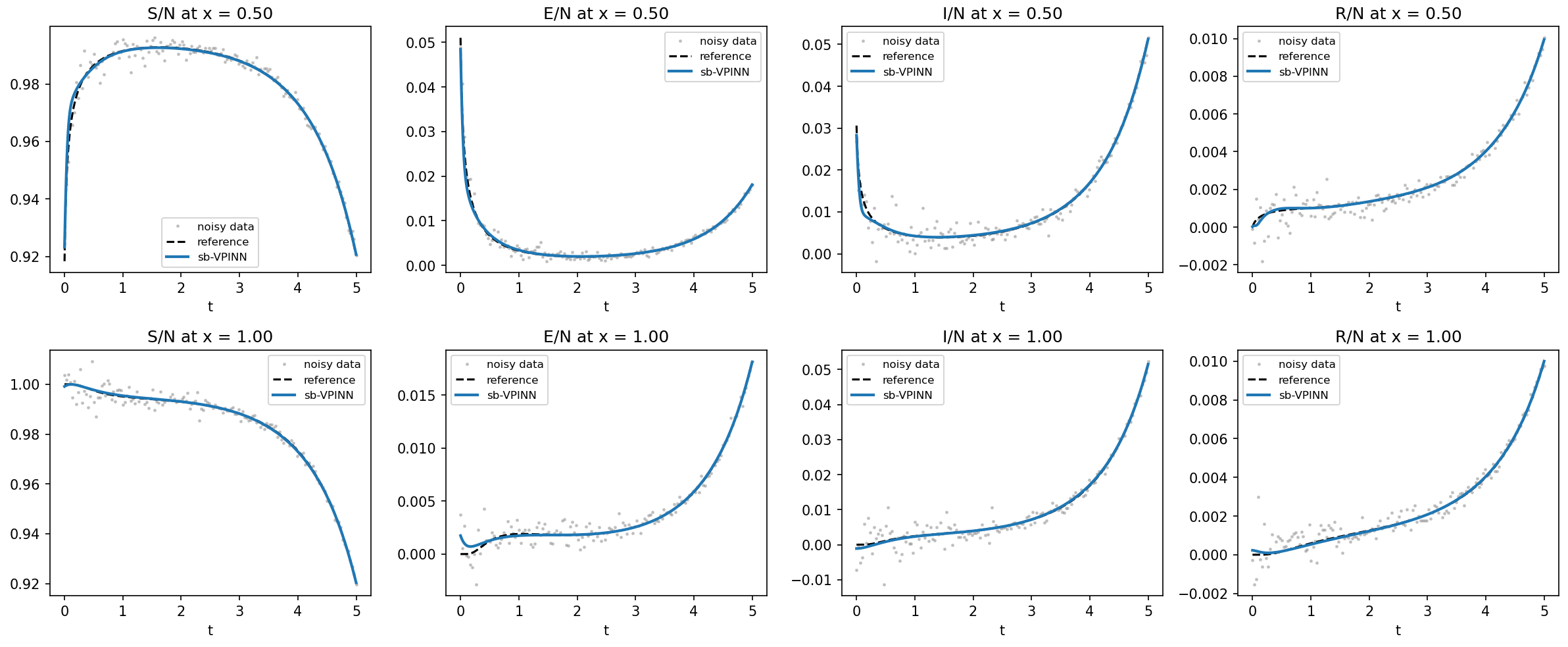}
\caption{Time-profile slices at $8\%$ observation noise, at two fixed spatial
locations, for the compartment proportions $U/N$. The learned fields track the
references; the noise appears as scatter that the network smooths through
rather than fits.}
\label{fig:profiles}
\end{figure}

For a representative run, Figure \ref{fig:spacetime} and Figure \ref{fig:profiles} illustrate the field reconstruction at $8\%$ observation noise, where the predicted compartment proportions track the clean reference curves at both spatial locations shown.

\subsection{The conditioning plane}
\label{sec:plane}

\begin{figure}[ht]
\centering
\includegraphics[width=0.6\linewidth]{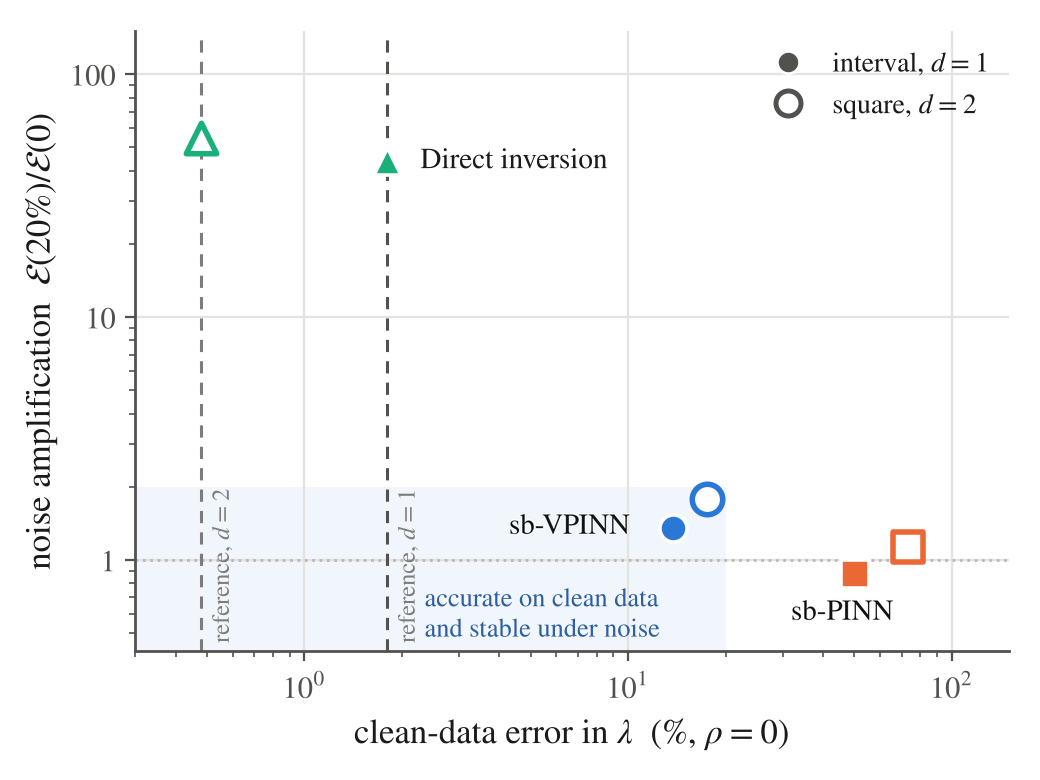}
\caption{The conditioning plane. Horizontal axis: mean clean-data error in $\lambda$; vertical axis: noise amplification $\Efun(20\%)/\Efun(0)$, the ratio of the mean error in $\lambda$ at $\rho=20\%$ to that on clean data. Filled markers:interval ($d=1$, Table \ref{tab:noise}); open markers: square ($d=2$, Table \ref{tab:2d}). Triangles: direct weak inversion; squares: sb-PINN (strong form); circles: sb-VPINN (weak form). Dashed lines mark the clean-field references of the direct inversion ($1.80\%$ for $d=1$, $0.48\%$ for $d=2$); the dotted line marks no amplification. The shaded region contains estimators whose clean-data error is at most $20\%$ and whose error at most doubles at $\rho=20\%$; in both dimensions only the weak form lies in it.}
\label{fig:plane}
\end{figure}

Difficulties D2 and D3 pose two different questions, and each baseline answers
only one. Figure \ref{fig:plane} places the three estimators on both axes at
once: clean-data error in $\lambda$ horizontally, noise amplification
$\Efun(20\%)/\Efun(0)$ vertically. The direct weak inversion sits at
$(1.80\%,43.9)$, the scale-balanced strong form at $(50.35\%,0.87)$ and the
scale-balanced weak form at $(13.72\%,1.35)$: the weak form is $3.7$ times more accurate than the strong form on clean data and amplifies noise about thirty times less than the direct inversion.

Each of the two usual protocols admits a degenerate pass. An estimator that is
flat in noise looks robust until one sees that it was already wrong at $\rho=0$, i.e., the strong form's $\lambda$ error barely moves with noise because it is near $50\%$ from the start. An estimator that is exact on clean data looks correct until noise is added: the direct inversion's $\lambda$ error rises from $1.80\%$ to $79.02\%$. Both axes must therefore be reported. Since the vertical axis is a ratio, we also record the absolute errors at $\rho=20\%$: $79.02\%$ for the direct inversion, $43.87\%$ for the strong form and $18.55\%$ for the weak form, which is thus the most accurate of the three in $\lambda$ at the largest noise level as well. The direct inversion of the clean field supplies the clean-field reference against which the horizontal axis is analyzed. It assembles its system from the complete observed grid, $200$ time levels of all $101$ nodes, three to five times as many space-time points as the neural estimators fit. Section \ref{sec:2d} shows that on the denser two-dimensional grid, its behaviour under noise changes.

\subsection{Two-dimensional validation}\label{sec:2d}
 
We repeat the three-estimator comparison on the square $\Omega_2=(0,L)^2$,
where $\Delta=\partial_{xx}+\partial_{yy}$ and the Neumann
condition \eqref{eq:neumann} holds on all four sides. The square-domain experiments retain the scale-balancing strategy and staged optimisation of Section \ref{sec:method}, while extending the network input to $(t,x,y)$ and using the point counts, test modes and training budgets specified below.

\paragraph{Test functions and weak residual.}
The one-dimensional cosine modes are replaced by their tensor products,
\begin{equation}
\varphi_{mn}(x,y)=\cos\Big(\frac{m\pi x}{L}\Big)\cos\Big(\frac{n\pi y}{L}\Big),
\qquad m,n=0,1,\dots,M_\varphi-1,
\label{eq:phi2d}
\end{equation}
which are the Neumann eigenfunctions of the Laplacian on the square. Green's
formula gives, for $u\in H^2(\Omega_2)$ with $\partial_n u=0$ on
$\partial\Omega_2$,
\begin{equation}
\begin{aligned}
-\lambda\int_{\Omega_2}\Delta u\varphi_{mn}\mathrm{d}z
&=-\lambda\int_{\partial\Omega_2}\varphi_{mn}\partial_n u\mathrm{d}s
+\lambda\int_{\Omega_2}\nabla u\cdot\nabla\varphi_{mn}\mathrm{d}z\\
&=\lambda\int_{\Omega_2}\big(u_x\partial_x\varphi_{mn}
+u_y\partial_y\varphi_{mn}\big)\mathrm{d}z ,
\end{aligned}
\label{eq:ibp2d}
\end{equation}
the boundary integral vanishing by \eqref{eq:neumann} exactly as in
Proposition \ref{prop:ibp}. The two-dimensional weak residuals are
therefore \eqref{eq:weakres} with $\langle f\rangle_m$ replaced by
$\langle f\rangle_{mn}=\int_{\Omega_2}f\varphi_{mn}\mathrm{d}z$ and each
diffusion term $\lambda\langle U_{\Theta,x},\varphi_m'\rangle$ replaced by
$\lambda\langle\nabla U_\Theta,\nabla\varphi_{mn}\rangle$. The network is
differentiated once in each variable, giving $U_{\Theta,t}$, $U_{\Theta,x}$
and $U_{\Theta,y}$, and never twice. The weak loss is \eqref{eq:weakloss} with
the mode sum running over the $M_\varphi^2$ pairs $(m,n)$ and the normalisation
$1/(N_fM_\varphi)$ replaced by $1/(N_fM_\varphi^2)$. The constant mode
$\varphi_{00}\equiv1$ is retained, as in one dimension. 

\paragraph{Setup.}
Reference data are generated by the finite-element solver of
Section \ref{sec:data} on the triangulated square, with the initial condition
of that section, and stored on a uniform tensor grid.
% TODO: state the grid size (N_x x N_y), N_t, L and the outbreak site z_c
%       used for the square dataset.
Noise follows \eqref{eq:noise} at the same six levels
$\rho\in\{0,4,8,12,16,20\}\%$ and seeds $0$-$4$. For a given seed a single
standard-normal field $\xi$ is drawn and scaled by $\rho$, so the six noise
levels of a seed share one noise pattern at six amplitudes. The scales $s_c$
are computed from the observed field by \eqref{eq:scale}, the sum now running
over the full space-time grid.
 
Both neural estimators use the $\tanh$ MLP of depth $6$ and width $64$ with
input $(t,x,y)$, each coordinate mapped affinely to $[-1,1]$, the output
non-dimensionalisation \eqref{eq:outscale}, the loss-side
weighting \eqref{eq:lossscale}, the softplus parameterisation and the
off-truth initialisation. Both fit the same $N_d=8000$ randomly sampled data points and the observed $t=0$ slice, and both train by
Algorithm \ref{alg:train} with $1500$ warm-up steps, $9000$ joint Adam steps and at most $300$ L-BFGS iterations. The weak form imposes the physics through $M_\varphi=6$ modes per axis ($36$ modes) at $N_f=40$ time levels, each evaluated on the full spatial grid, with the constraint terms on $N_c=8000$ random points and no boundary set. The strong form imposes the pointwise residual with the automatic-differentiation Laplacian $U_{\Theta,xx}+U_{\Theta,yy}$ at $8000$ random collocation points, which also carry the constraint terms, and adds a Neumann penalty on $120$ random points per edge. The direct weak inversion of Section \ref{sec:direct} is assembled from $M_\varphi=8$ modes per axis ($64$ modes) and $200$ time levels, with finite-difference derivatives. At $\rho=0$ it acts on the exact field, and for $\rho>0$ on the complete observed grid.
 
These settings differ from those of the one-dimensional runs ($M_\varphi=8$ and
$N_f=80$ in the weak residual, $1000$ warm-up and $7000$ Adam steps), because
the cost of the weak residual grows with $N_f$ times the number of spatial
nodes. %The comparison between the three estimators within each dimension is matched; the comparison of each estimator across dimensions is not a controlled experiment in dimension alone.

\begin{figure}[htbp]
    \centering

    \begin{subfigure}[t]{0.49\textwidth}
        \centering
        \includegraphics[width=\linewidth]{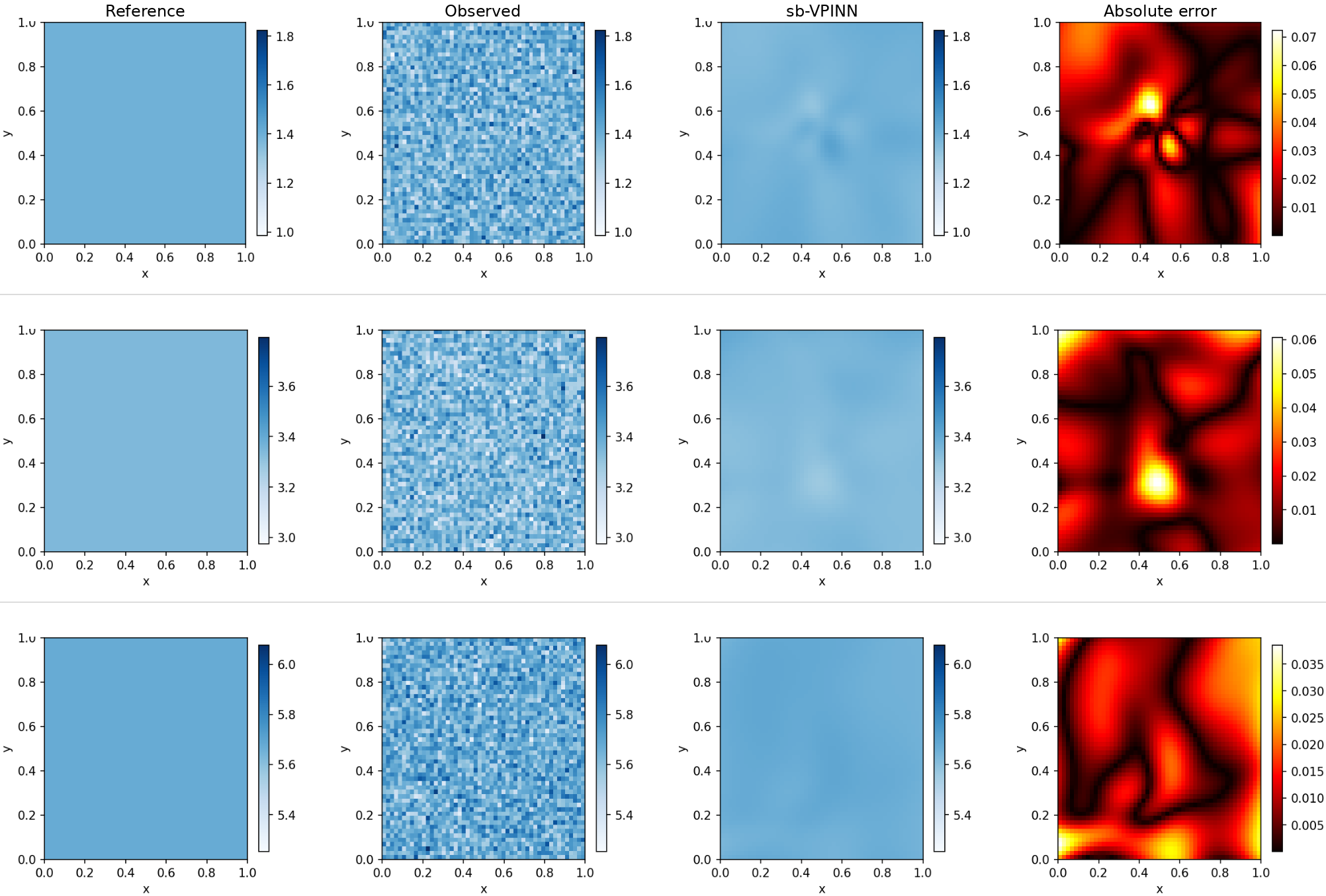}
        \caption{Susceptible population $S$.}
        \label{fig:snapshots-S}
    \end{subfigure}\hfill
    \begin{subfigure}[t]{0.49\textwidth}
        \centering
        \includegraphics[width=\linewidth]{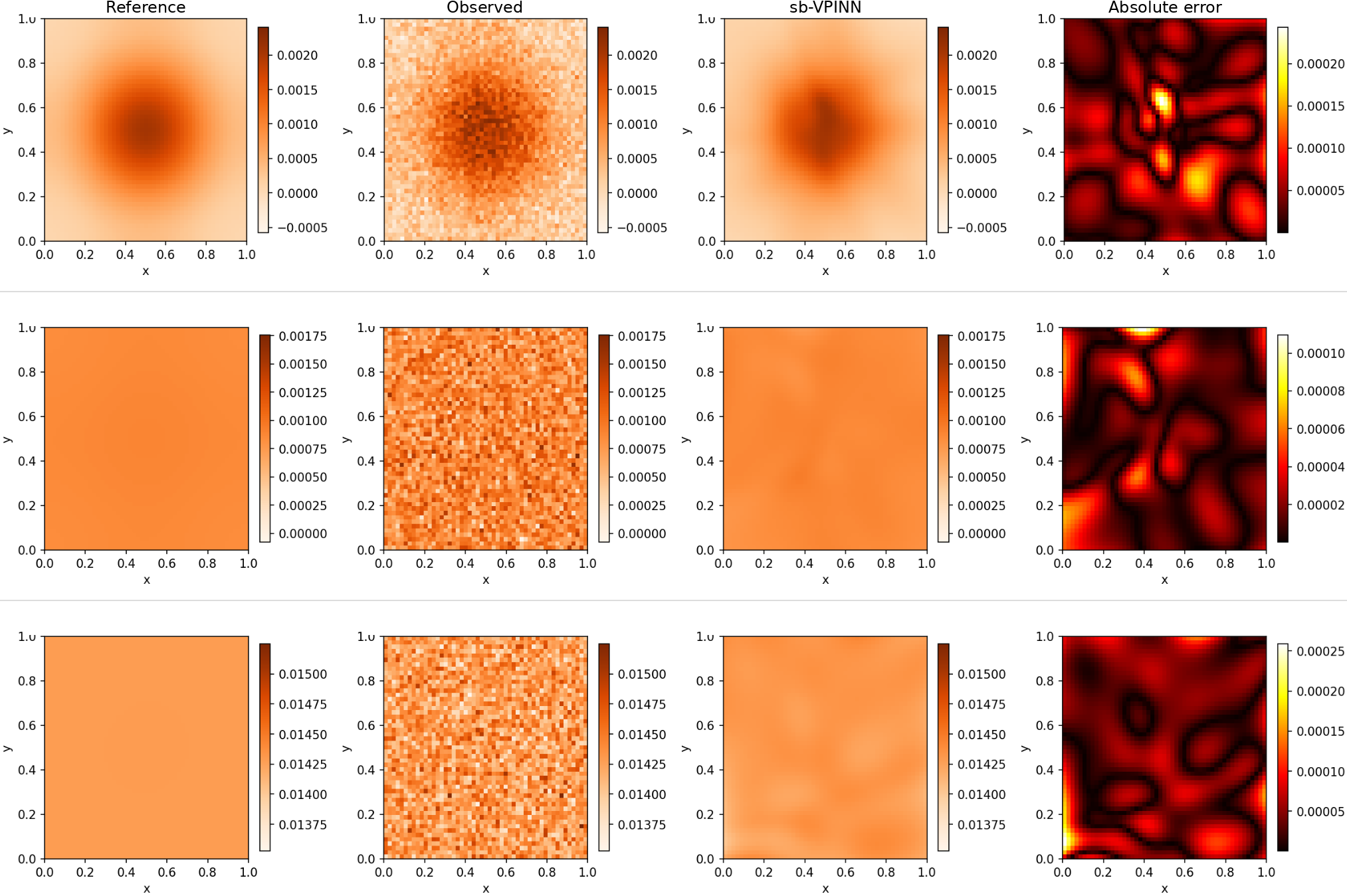}
        \caption{Exposed population $E$.}
        \label{fig:snapshots-E}
    \end{subfigure}

    \medskip

    \begin{subfigure}[t]{0.49\textwidth}
        \centering
        \includegraphics[width=\linewidth]{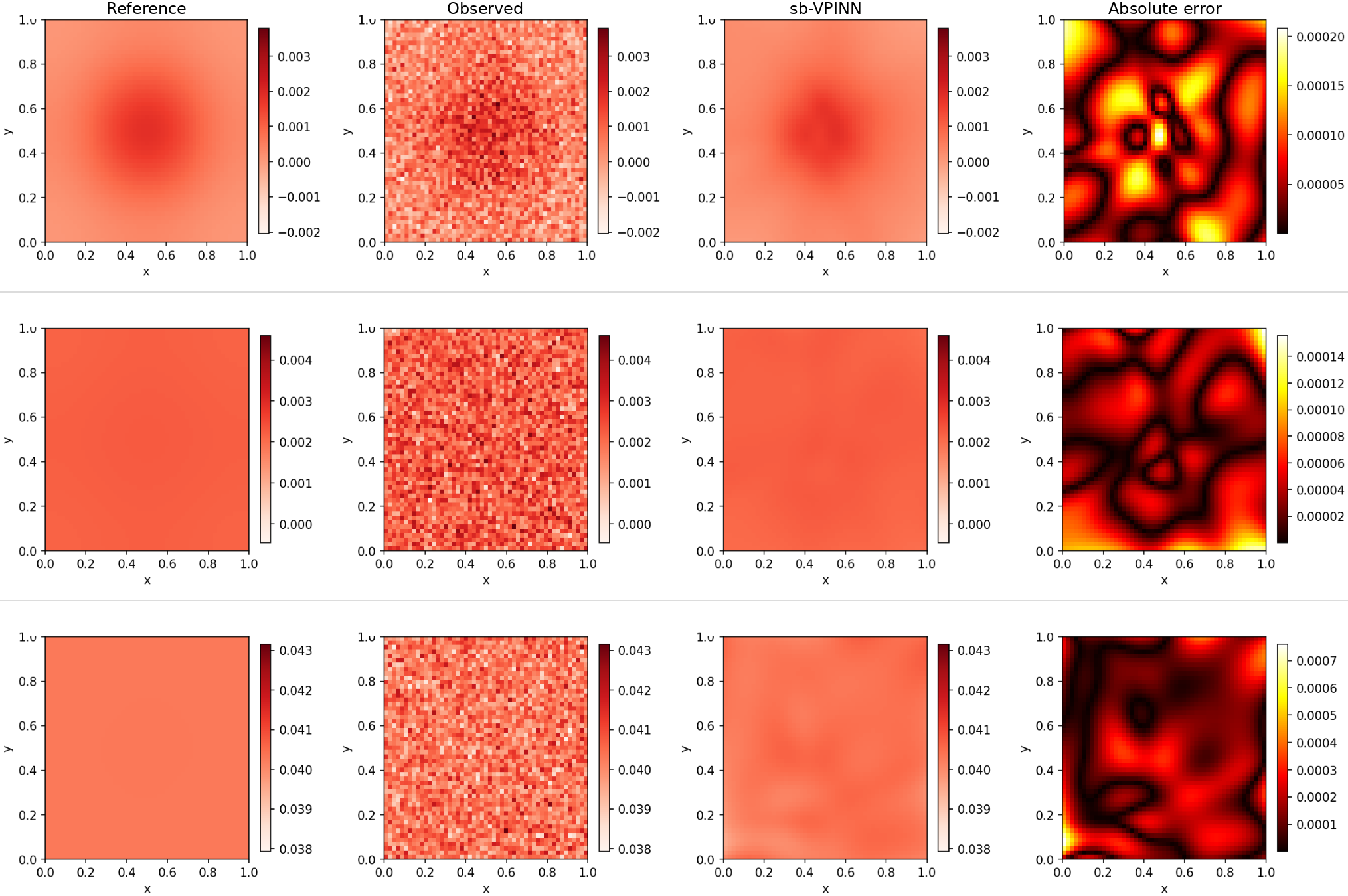}
        \caption{Infectious population $I$.}
        \label{fig:snapshots-I}
    \end{subfigure}\hfill
    \begin{subfigure}[t]{0.49\textwidth}
        \centering
        \includegraphics[width=\linewidth]{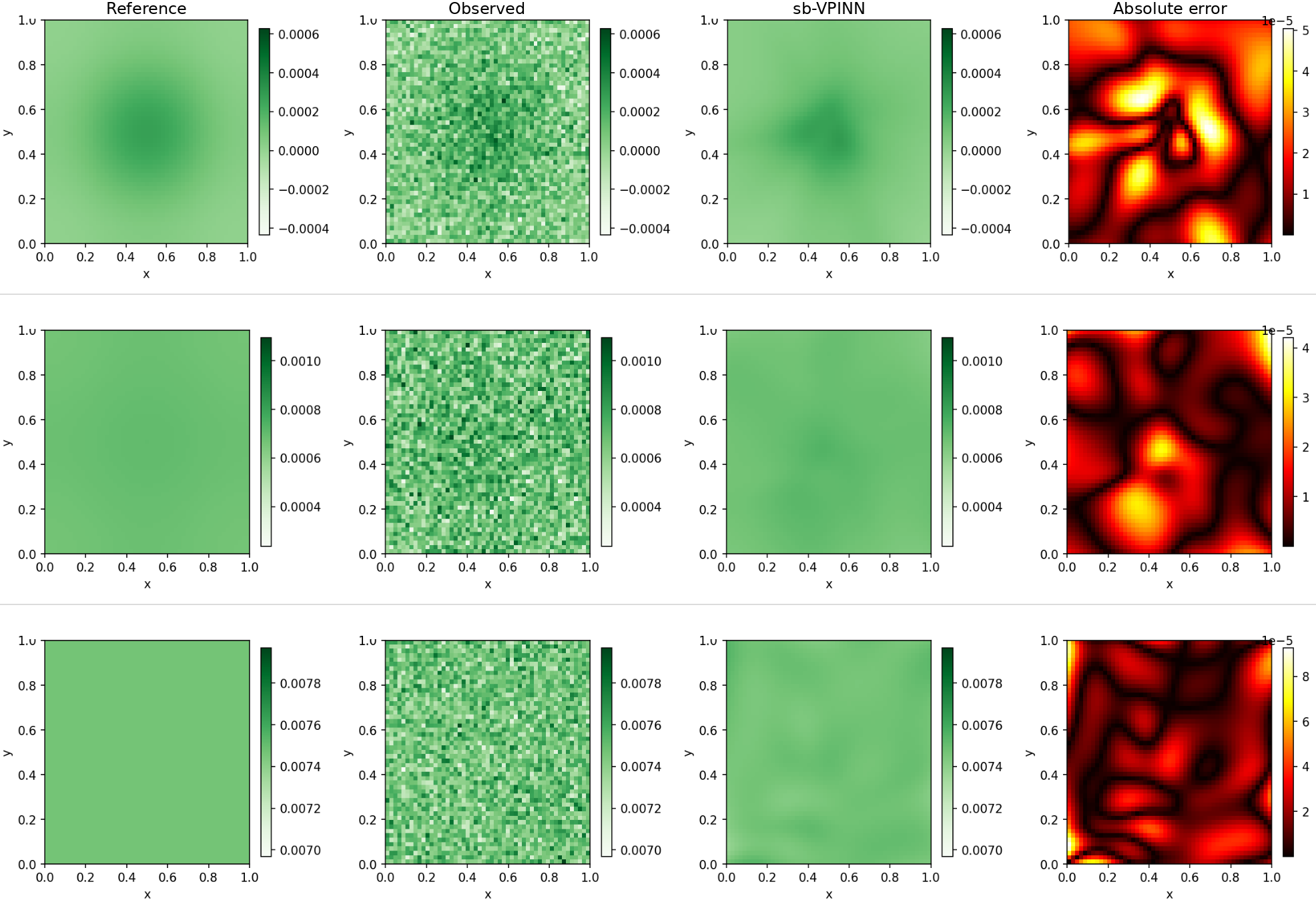}
        \caption{Recovered population $R$.}
        \label{fig:snapshots-R}
    \end{subfigure}

    \caption{Reconstruction of the SEIR fields from $8\%$ noisy observations using sb-VPINN. Panels (a)-(d) correspond to $S$, $E$, $I$, and $R$. Rows show $t=0.5$, $2.5$, and $5$ from top to bottom; columns show the reference field, noisy
observations, sb-VPINN prediction, and pointwise absolute error on $[0,1]^2$. Within each row, the first three columns share a colour scale; the error has a separate scale. Scales vary across populations and times, with errors of order $10^{-2}$ for $S$, $10^{-4}$ for $E$ and $I$, and $10^{-5}$ for $R$.}
    \label{fig:seir-snapshots}
\end{figure}

\begin{table}[ht]
\centering
\caption{Two-dimensional square: relative parameter errors (\%), mean $\pm$ sample standard deviation over seeds $0$-$4$, for the direct weak inversion, the scale-balanced strong-form estimator and the scale-balanced weak-form
estimator. The per-seed record is given in Table \ref{tab:perseed2d}.}
\label{tab:2d}
\footnotesize
\setlength{\tabcolsep}{4.5pt}
\resizebox{\linewidth}{!}{%
\begin{tabular}{llcccccc}
\toprule
Method & Param. & Clean & $\rho=4\%$ & $\rho=8\%$ & $\rho=12\%$ & $\rho=16\%$ & $\rho=20\%$\\
\midrule
\multirow{4}{*}{Direct weak inversion}
 & $\beta$   & $0.21\pm0.00$ & $0.71\pm0.41$ & $1.46\pm0.70$ & $2.23\pm1.02$ & $2.99\pm1.35$ & $3.76\pm1.68$\\
 & $\delta$  & $0.21\pm0.00$ & $4.93\pm3.99$ & $9.97\pm7.74$ & $15.06\pm11.42$ & $20.17\pm15.08$ & $25.27\pm18.74$\\
 & $\gamma$  & $0.22\pm0.00$ & $1.46\pm0.47$ & $2.88\pm1.16$ & $4.29\pm1.85$ & $5.71\pm2.56$ & $7.12\pm3.27$\\
 & $\lambda$ & $0.48\pm0.00$ & $1.19\pm1.24$ & $4.90\pm2.59$ & $10.87\pm3.94$ & $18.21\pm5.20$ & $26.06\pm6.30$\\
\midrule
\multirow{4}{*}{sb-PINN}
 & $\beta$   & $0.34\pm0.15$ & $0.27\pm0.15$ & $0.27\pm0.17$ & $0.21\pm0.14$ & $0.26\pm0.14$ & $0.22\pm0.21$\\
 & $\delta$  & $3.91\pm0.65$ & $5.50\pm2.84$ & $6.07\pm2.13$ & $5.52\pm1.90$ & $6.71\pm1.66$ & $5.45\pm2.34$\\
 & $\gamma$  & $0.22\pm0.12$ & $0.24\pm0.17$ & $0.25\pm0.19$ & $0.43\pm0.13$ & $0.43\pm0.16$ & $0.41\pm0.17$\\
 & $\lambda$ & $73.12\pm16.52$ & $78.20\pm6.75$ & $79.04\pm8.08$ & $79.90\pm7.71$ & $83.18\pm6.26$ & $82.26\pm4.30$\\
\midrule
\multirow{4}{*}{\textbf{sb-VPINN}}
 & $\beta$   & $0.23\pm0.04$ & $0.23\pm0.04$ & $0.22\pm0.12$ & $0.24\pm0.15$ & $0.27\pm0.23$ & $0.36\pm0.22$\\
 & $\delta$  & $0.68\pm0.33$ & $0.53\pm0.35$ & $0.57\pm0.51$ & $0.81\pm0.65$ & $1.12\pm0.77$ & $1.42\pm1.14$\\
 & $\gamma$  & $0.18\pm0.02$ & $0.17\pm0.11$ & $0.20\pm0.15$ & $0.21\pm0.25$ & $0.33\pm0.43$ & $0.51\pm0.46$\\
 & $\lambda$ & $17.51\pm5.67$ & $23.96\pm7.59$ & $25.39\pm8.05$ & $26.93\pm5.97$ & $25.80\pm8.58$ & $31.15\pm7.49$\\
\bottomrule
\end{tabular}}
\end{table}

\paragraph{sb-VPINN compared against sb-PINN (in 2D).}
Table \ref{tab:2d} aggregates the three estimators over five seeds and six noise levels, and Table \ref{tab:perseed2d} gives the per-seed record. The matched comparison is more decisive than in one dimension. The weak form
attains the lower $\lambda$ error in every one of the $30$ runs, for every seed at every noise level. The strong-form diffusion error is $73.12\pm16.52\%$ on clean data, with seed means ranging from $78.20\%$ to $83.18\%$ at nonzero noise
levels. The corresponding weak-form values are $17.51\pm5.67\%$ on clean data and seed means between $23.96\%$ and $31.15\%$ under noise. At every noise level, its mean diffusion error is between $2.64$ and $4.18$ times smaller than that of the strong form. In two dimensions, the strong form also loses accuracy in a reaction parameter. Its error in $\delta$ is $3.9$-$6.7\%$ at every noise level, and the weak form's is lower in all $30$ runs. The seed-mean errors in $\beta$ and $\gamma$ remain below $0.7\%$
for both neural estimators.

\paragraph{The conditioning plane in two dimensions.}
As in Section \ref{sec:plane}, we compare the mean error in $\lambda$ on clean data with the ratio of its mean error at $\rho=20\%$ to that on clean data. The matched comparison is decisive. The strong form's diffusion error is already
$73.12\%$ on clean data and $82.26\%$ at $\rho=20\%$. The weak form reduces these to $17.51\%$ and $31.15\%$, and its $\lambda$ error is lower in every one of the $30$ runs, for every seed at every noise level, by a factor of $3.3$ on average (geometric mean). It does so without giving up the reaction rates: $\beta$, $\delta$ and $\gamma$ stay below $1.5\%$ throughout, and its error in $\delta$ is also lower than the strong form's in all $30$ runs. Between clean data and $\rho=20\%$ the weak form's $\lambda$ error grows by a factor of $1.78$.

The direct inversion works from different data and answers a different question. It assembles its linear system from the complete observed grid, all $2601$ nodes of all four compartments at $200$ time levels, about $5.2\times10^5$ space-time points, whereas both neural estimators fit $8000$ sampled points and the initial slice, about $1.1\times10^4$: roughly fifty times fewer. On the exact field it recovers $\lambda$ to $0.48\%$, which shows that the weak system determines $\lambda$ accurately when the full clean field is known and that is taken as the clean-field reference on the square. Under noise its $\lambda$ error grows roughly quadratically, by a factor of $54.3$ over the sweep, yet remains below the weak form's at every noise level ($26.06\%$ against $31.15\%$ at $\rho=20\%$), while its reaction rates degrade linearly to $3.76\%$, $25.27\%$ and $7.12\%$ in $\beta$, $\delta$ and $\gamma$, ten to eighteen times the weak form's errors. It is therefore a reference for what the weak system yields when every grid value is observed, a regime that epidemiological surveillance does not provide, and not a competitor to estimators that work from scattered samples.

In summary, the scale-balanced weak form carries over to the square with no change beyond the tensor-product test space \eqref{eq:phi2d}, and its advantage over the matched strong form is more consistent than in one dimension: a lower
$\lambda$ error in all $30$ runs rather than $28$ of $30$, about threefold on average, together with a lower error in $\delta$. Among the three estimators it is the only one that works from scattered samples and keeps all three reaction
rates below $1.5\%$ at every noise level. The direct inversion recovers $\lambda$ to $0.48\%$ from the clean field and, given roughly fifty times more observations, recovers it more accurately under spatially independent noise, at the cost of an order of magnitude in the reaction rates.
%Irregular and unstructured geometries, where the weak form's mesh-native first-derivative assembly becomes essential and a finite-difference Laplacian no longer applies, are deferred to the companion study.

% =============================================================================
\section{Discussion and conclusion}
% =============================================================================

The sb-VPINN lowers the error in $\lambda$ relative to the matched scale-balanced strong form (Q1) while leaving $\beta,\delta,\gamma$ near the floor (Q2): it is a targeted repair of one failure mode rather than a uniform accuracy improvement. Under noise the four components of $\theta$ degrade at very different rates, $\lambda$ is the dominant fragile direction, and the weak form retains its advantage across the sweep (Q3). Direct weak inversion provides a clean-field numerical reference and a separate assessment of sensitivity to observation noise.

What we regard as most transferable is the combination of the two ingredients, which are not interchangeable. Scale balancing repairs the reaction parameters and does nothing for the diffusion coefficient essentially. The weak residual repairs the diffusion coefficient but, used alone, would leave the reaction parameters buried under the susceptible field by \eqref{eq:imbalance}. Each addresses a difficulty the other cannot reach, and in a coupled, scale-separated coefficient inverse problem both are needed.

Three limitations bound the claims. First, diffusion recovery remains imperfect: across the tested noise levels, the weak-form mean error in $\lambda$ ranges from $13.72\%$ to $18.55\%$ in one dimension and from $17.51\%$ to $31.15\%$ in two dimensions. Secondly, we tested independently generated Gaussian noise. Errors that persist over time or affect neighbouring locations together were not considered. Finally, the reference fields are numerical solutions of the assumed model, so the experiments establish parameter-recovery performance within that model, while validation against real epidemic data remains necessary.

In a spatial epidemic model the reaction rates describe how quickly the disease runs through the population at a given place, whereas the diffusivity $\lambda$ describes how quickly it moves across the map. In reaction-diffusion models of this type, the speed of the invading front grows with $\sqrt{\lambda}$ \citep{murray2003}. Questions such as when an outbreak will reach a neighbouring district, how wide a buffer or vaccination zone should be, or how much lead time a health authority has, therefore depend on $\lambda$ \citep{riley2007,viguerie2021}, and $\lambda$ is exactly the quantity a standard fit gets wrong. Reducing its mean error while keeping each reaction rate as accurate as possible makes a data-calibrated spatial model more useful for assessing spatial spread and understanding its dynamics. The same combination of compartments of very different size and a coefficient that acts only through curvature arises whenever a diffusion rate is estimated in a reaction-diffusion model in ecology or cell biology \citep{murray2003}, and both ingredients carry over unchanged.

Two natural curvature problem extensions follow from this work. On irregular, real-geography domains the weak form's mesh-native first-derivative assembly is not merely preferable but necessary, since a finite-difference Laplacian no longer applies the motivation for the companion study on unstructured meshes. And because \eqref{eq:direct} is a linear system in $\theta$ for a fixed field, it admits a Bayesian treatment returning credible intervals rather than point estimates \citep{kaipio2005}, which is what decision-relevant epidemiology requires. Interventions such as lockdowns or school closures make the transmission rate time-dependent, and PINNs have already recovered a time-varying contact rate in non-spatial compartmental models \citep{grimm2022}. Combining this with the weak-form treatment of $\lambda$ would allow the effect of an intervention and the speed of spatial spread to be estimated together. Human mobility is also far from uniform, so a spatially varying diffusivity $\lambda(x)$, informed for instance by commuting or transport networks, is the more realistic target, although recovering a coefficient that varies in space is a harder and less well-posed inverse problem than recovering a constant \citep{richter1981,isakov2017}.

\section*{Data and code availability}
The synthetic finite-element datasets, the trained-model outputs and the scripts that regenerate every figure and table in this article will be deposited in a public repository on acceptance and are available from the corresponding author in the meantime.

\section*{Reproducibility}

Each run is fixed by a seed $k\in\{0,\dots,4\}$, which determines the network initialisation, the observation noise and the point sampling. The noise is drawn on the full stored grid before any point is sampled, so for a given dataset,
seed and noise level the noisy field does not depend on the point counts. Since the generator is reseeded for every run, the six noise levels of a seed share a single noise pattern across six amplitudes. The sampled points and the observation-derived
scales $s_c$ are fixed before training and remain unchanged throughout. All computations use 64-bit floating point, with PyTorch's deterministic algorithms enabled, cuDNN benchmarking disabled and \texttt{CUBLAS\_WORKSPACE\_CONFIG=:4096:8} set before PyTorch is imported; as this variable takes effect only if PyTorch has not already been loaded, each run is started in a fresh process. The direct inversions run in CPU NumPy and are deterministic. All experiments were run on the \texttt{v100} server of the Institut de Recherche Math\'ematique Avanc\'ee (IRMA), Universit\'e de
Strasbourg, a virtual machine with 48 AMD CPU cores and one NVIDIA V100 GPU. The neural estimators were trained on the GPU and the direct inversions were run on the CPU, with Python 3.10.20, PyTorch 2.4.0 (CUDA 11.8), NumPy 2.2.6 and SciPy 1.15.2. Reproduction requires the same dataset, effective configuration, seed, and software and hardware environment. Bitwise agreement across different environments is not assumed. Hence, none of our conclusions rests on a single run, and every reported error is the mean and sample standard deviation over five seeds, and comparisons are tested per noise level across seeds.

\section*{Competing interests}
The authors have no relevant financial or non-financial interests to disclose.

\section*{Author contributions}

A.B.: conceptualisation, methodology, formal analysis, writing - original draft and review. S.S.: conceptualisation, software, investigation, simulation, validation, writing - original draft and review. 

\section*{Acknowledgements}
Computations were performed on the server of the Institut de
Recherche Math\'ematique Avanc\'ee (IRMA), Universit\'e de Strasbourg. The authors thank the IRMA computing team for access and technical support.

\bibliographystyle{unsrtnat}
\bibliography{references}

@article{kermack1927,
  author  = {Kermack, William Ogilvy and McKendrick, Anderson Gray},
  title   = {A Contribution to the Mathematical Theory of Epidemics},
  journal = {Proceedings of the Royal Society of London. Series A},
  volume  = {115}, number = {772}, pages = {700--721}, year = {1927},
  doi     = {10.1098/rspa.1927.0118}
}

@article{hethcote2000,
  author  = {Hethcote, Herbert W.},
  title   = {The Mathematics of Infectious Diseases},
  journal = {SIAM Review},
  volume  = {42}, number = {4}, pages = {599--653}, year = {2000},
  doi     = {10.1137/S0036144500371907}
}

@article{vandendriessche2002,
  author  = {van den Driessche, P. and Watmough, James},
  title   = {Reproduction Numbers and Sub-threshold Endemic Equilibria for
             Compartmental Models of Disease Transmission},
  journal = {Mathematical Biosciences},
  volume  = {180}, number = {1--2}, pages = {29--48}, year = {2002},
  doi     = {10.1016/S0025-5564(02)00108-6}
}

@book{murray2003,
  author    = {Murray, James D.},
  title     = {Mathematical Biology {II}: Spatial Models and Biomedical Applications},
  series    = {Interdisciplinary Applied Mathematics}, volume = {18},
  edition   = {3rd}, publisher = {Springer-Verlag}, address = {New York},
  year      = {2003}, doi = {10.1007/b98869}
}

@article{riley2007,
  author  = {Riley, Steven},
  title   = {Large-Scale Spatial-Transmission Models of Infectious Disease},
  journal = {Science},
  volume  = {316}, number = {5829}, pages = {1298--1301}, year = {2007},
  doi     = {10.1126/science.1134695}
}

@article{viguerie2021,
  author  = {Viguerie, Alex and Lorenzo, Guillermo and Auricchio, Ferdinando and
             Baroli, Davide and Hughes, Thomas J. R. and Patton, Alessia and
             Reali, Alessandro and Yankeelov, Thomas E. and Veneziani, Alessandro},
  title   = {Simulating the Spread of {COVID-19} via a Spatially-Resolved
             Susceptible--Exposed--Infected--Recovered--Deceased ({SEIRD}) Model
             with Heterogeneous Diffusion},
  journal = {Applied Mathematics Letters},
  volume  = {111}, pages = {106617}, year = {2021},
  doi     = {10.1016/j.aml.2020.106617}
}

@article{parshad2023,
  author  = {Parshad, Rana D. and Wickramasooriya, Sureni and
             Antwi-Fordjour, Kwadwo and Banerjee, Aniket},
  title   = {Additional Food Causes Predators to Explode --- Unless the
             Predators Compete},
  journal = {International Journal of Bifurcation and Chaos},
  volume  = {33}, number = {3}, pages = {2350034}, year = {2023},
  doi     = {10.1142/S0218127423500347}
}

@article{raissi2019,
  author  = {Raissi, Maziar and Perdikaris, Paris and Karniadakis, George Em},
  title   = {Physics-Informed Neural Networks: A Deep Learning Framework for
             Solving Forward and Inverse Problems Involving Nonlinear Partial
             Differential Equations},
  journal = {Journal of Computational Physics},
  volume  = {378}, pages = {686--707}, year = {2019},
  doi     = {10.1016/j.jcp.2018.10.045}
}

@article{karniadakis2021,
  author  = {Karniadakis, George Em and Kevrekidis, Ioannis G. and Lu, Lu and
             Perdikaris, Paris and Wang, Sifan and Yang, Liu},
  title   = {Physics-Informed Machine Learning},
  journal = {Nature Reviews Physics},
  volume  = {3}, number = {6}, pages = {422--440}, year = {2021},
  doi     = {10.1038/s42254-021-00314-5}
}

@article{cuomo2022,
  author  = {Cuomo, Salvatore and Schiano Di Cola, Vincenzo and Giampaolo, Fabio
             and Rozza, Gianluigi and Raissi, Maziar and Piccialli, Francesco},
  title   = {Scientific Machine Learning Through Physics-Informed Neural
             Networks: Where We Are and What's Next},
  journal = {Journal of Scientific Computing},
  volume  = {92}, number = {3}, pages = {88}, year = {2022},
  doi     = {10.1007/s10915-022-01939-z}
}

@article{lu2021deepxde,
  author  = {Lu, Lu and Meng, Xuhui and Mao, Zhiping and Karniadakis, George Em},
  title   = {{DeepXDE}: A Deep Learning Library for Solving Differential Equations},
  journal = {SIAM Review},
  volume  = {63}, number = {1}, pages = {208--228}, year = {2021},
  doi     = {10.1137/19M1274067}
}

@article{wang2021,
  author  = {Wang, Sifan and Teng, Yujun and Perdikaris, Paris},
  title   = {Understanding and Mitigating Gradient Flow Pathologies in
             Physics-Informed Neural Networks},
  journal = {SIAM Journal on Scientific Computing},
  volume  = {43}, number = {5}, pages = {A3055--A3081}, year = {2021},
  doi     = {10.1137/20M1318043}
}

@article{wang2022ntk,
  author  = {Wang, Sifan and Yu, Xinling and Perdikaris, Paris},
  title   = {When and Why {PINNs} Fail to Train: A Neural Tangent Kernel Perspective},
  journal = {Journal of Computational Physics},
  volume  = {449}, pages = {110768}, year = {2022},
  doi     = {10.1016/j.jcp.2021.110768}
}

@inproceedings{krishnapriyan2021,
  author    = {Krishnapriyan, Aditi S. and Gholami, Amir and Zhe, Shandian and
               Kirby, Robert M. and Mahoney, Michael W.},
  title     = {Characterizing Possible Failure Modes in Physics-Informed
               Neural Networks},
  booktitle = {Advances in Neural Information Processing Systems 34
               ({NeurIPS} 2021)},
  pages     = {26548--26560}, year = {2021},
  note      = {arXiv:2109.01050}
}

@article{kharazmi2019vpinn,
  author  = {Kharazmi, Ehsan and Zhang, Zhongqiang and Karniadakis, George Em},
  title   = {Variational Physics-Informed Neural Networks for Solving Partial
             Differential Equations},
  journal = {arXiv preprint arXiv:1912.00873}, year = {2019},
  doi     = {10.48550/arXiv.1912.00873}
}

@article{kharazmi2021hpvpinn,
  author  = {Kharazmi, Ehsan and Zhang, Zhongqiang and Karniadakis, George Em},
  title   = {{hp-VPINNs}: Variational Physics-Informed Neural Networks with
             Domain Decomposition},
  journal = {Computer Methods in Applied Mechanics and Engineering},
  volume  = {374}, pages = {113547}, year = {2021},
  doi     = {10.1016/j.cma.2020.113547}
}

@article{weinan2018deepritz,
  author  = {E, Weinan and Yu, Bing},
  title   = {The Deep {Ritz} Method: A Deep Learning-Based Numerical Algorithm
             for Solving Variational Problems},
  journal = {Communications in Mathematics and Statistics},
  volume  = {6}, number = {1}, pages = {1--12}, year = {2018},
  doi     = {10.1007/s40304-018-0127-z}
}

@article{zang2020wan,
  author  = {Zang, Yaohua and Bao, Gang and Ye, Xiaojing and Zhou, Haomin},
  title   = {Weak Adversarial Networks for High-Dimensional Partial
             Differential Equations},
  journal = {Journal of Computational Physics},
  volume  = {411}, pages = {109409}, year = {2020},
  doi     = {10.1016/j.jcp.2020.109409}
}

@article{bao2020wan,
  author  = {Bao, Gang and Ye, Xiaojing and Zang, Yaohua and Zhou, Haomin},
  title   = {Numerical Solution of Inverse Problems by Weak Adversarial Networks},
  journal = {Inverse Problems},
  volume  = {36}, number = {11}, pages = {115003}, year = {2020},
  doi     = {10.1088/1361-6420/abb447}
}

@article{brunton2016,
  author  = {Brunton, Steven L. and Proctor, Joshua L. and Kutz, J. Nathan},
  title   = {Discovering Governing Equations from Data by Sparse Identification
             of Nonlinear Dynamical Systems},
  journal = {Proceedings of the National Academy of Sciences},
  volume  = {113}, number = {15}, pages = {3932--3937}, year = {2016},
  doi     = {10.1073/pnas.1517384113}
}

@article{schaeffer2017,
  author  = {Schaeffer, Hayden and McCalla, Scott G.},
  title   = {Sparse Model Selection via Integral Terms},
  journal = {Physical Review E},
  volume  = {96}, number = {2}, pages = {023302}, year = {2017},
  doi     = {10.1103/PhysRevE.96.023302}
}

@article{reinbold2020,
  author  = {Reinbold, Patrick A. K. and Gurevich, Daniel R. and
             Grigoriev, Roman O.},
  title   = {Using Noisy or Incomplete Data to Discover Models of
             Spatiotemporal Dynamics},
  journal = {Physical Review E},
  volume  = {101}, number = {1}, pages = {010203(R)}, year = {2020},
  doi     = {10.1103/PhysRevE.101.010203}
}

@article{messenger2021,
  author  = {Messenger, Daniel A. and Bortz, David M.},
  title   = {Weak {SINDy} for Partial Differential Equations},
  journal = {Journal of Computational Physics},
  volume  = {443}, pages = {110525}, year = {2021},
  doi     = {10.1016/j.jcp.2021.110525}
}

@article{kharazmi2021nat,
  author  = {Kharazmi, Ehsan and Cai, Min and Zheng, Xiaoning and Zhang, Zhen and
             Lin, Guang and Karniadakis, George Em},
  title   = {Identifiability and Predictability of Integer- and Fractional-Order
             Epidemiological Models Using Physics-Informed Neural Networks},
  journal = {Nature Computational Science},
  volume  = {1}, number = {11}, pages = {744--753}, year = {2021},
  doi     = {10.1038/s43588-021-00158-0}
}

@article{grimm2022,
  author  = {Grimm, Viktor and Heinlein, Alexander and Klawonn, Axel and
             Lanser, Martin and Weber, Janine},
  title   = {Estimating the Time-Dependent Contact Rate of {SIR} and {SEIR}
             Models in Mathematical Epidemiology Using Physics-Informed
             Neural Networks},
  journal = {Electronic Transactions on Numerical Analysis},
  volume  = {56}, pages = {1--27}, year = {2022},
  doi     = {10.1553/etna_vol56s1}
}

@article{millevoi2024,
  author  = {Millevoi, Caterina and Pasetto, Damiano and Ferronato, Massimiliano},
  title   = {A Physics-Informed Neural Network Approach for Compartmental
             Epidemiological Models},
  journal = {PLOS Computational Biology},
  volume  = {20}, number = {9}, pages = {e1012387}, year = {2024},
  doi     = {10.1371/journal.pcbi.1012387}
}

@article{zinihi2026,
  author  = {Zinihi, Achraf and Ehrhardt, Matthias},
  title   = {Constraint-Aware Physics-Informed Neural Networks for {SEIR}
             Reaction--Diffusion Epidemic Models with Vital Dynamics},
  journal = {arXiv preprint arXiv:2605.19886}, year = {2026},
  doi     = {10.48550/arXiv.2605.19886}
}

@article{richter1981,
  author  = {Richter, Gerard R.},
  title   = {An Inverse Problem for the Steady State Diffusion Equation},
  journal = {SIAM Journal on Applied Mathematics},
  volume  = {41}, number = {2}, pages = {210--221}, year = {1981},
  doi     = {10.1137/0141016}
}

@book{isakov2017,
  author    = {Isakov, Victor},
  title     = {Inverse Problems for Partial Differential Equations},
  series    = {Applied Mathematical Sciences}, volume = {127}, edition = {3rd},
  publisher = {Springer}, address = {Cham}, year = {2017},
  doi       = {10.1007/978-3-319-51658-5}
}

@book{engl1996,
  author    = {Engl, Heinz W. and Hanke, Martin and Neubauer, Andreas},
  title     = {Regularization of Inverse Problems},
  series    = {Mathematics and Its Applications}, volume = {375},
  publisher = {Kluwer Academic Publishers}, address = {Dordrecht}, year = {1996},
  isbn      = {978-0-7923-4157-4}
}

@book{kaipio2005,
  author    = {Kaipio, Jari P. and Somersalo, Erkki},
  title     = {Statistical and Computational Inverse Problems},
  series    = {Applied Mathematical Sciences}, volume = {160},
  publisher = {Springer}, address = {New York}, year = {2005},
  doi       = {10.1007/b138659}
}

@article{bansal2007,
     author  = {Bansal, Shweta and Grenfell, Bryan T. and Meyers, Lauren Ancel},
     title   = {When individual behaviour matters: homogeneous and network models in epidemiology},
     journal = {Journal of the Royal Society Interface},
     volume  = {4}, number = {16}, pages = {879--891}, year = {2007},
     doi     = {10.1098/rsif.2007.1100}
}

@article{rabiei2026ai,
  title={How AI Can Advance Mathematical Biology: Opportunities, Challenges, and Future Directions},
  author={Rabiei, Kobra and Rempala, Grzegorz A and Laubenbacher, Reinhard and Hao, Wenrui},
  journal={Bulletin of Mathematical Biology},
  volume={88},
  number={10},
  pages={175},
  year={2026},
  publisher={Springer}
}

\appendix

\section*{Appendix}

\section{Per-seed results for 1D}
\label{app:perseed}

Table \ref{tab:perseed} lists the relative error of every iterations. Table \ref{tab:noise} gives their means and sample standard deviations. The columns Dir., sf and wf are the direct weak inversion, the scale-balanced strong form, and the scale-balanced weak form, respectively. For a given seed, the six noise levels share one noise pattern at varying amplitude. Thus, entries in a column are correlated across noise levels. It is observed that direct inversion is much more variable across seeds under noise as it varies, while seed change does not affect the other control study (scale-balanced strong form) much.

\begin{table}[H]
\centering
\caption{Per-seed relative parameter errors (\%) for the direct weak inversion
(Dir.), the scale-balanced strong-form estimator (sf) and the scale-balanced
weak-form estimator (wf), across five seeds and six noise levels.}
\label{tab:perseed}
\tiny
\setlength{\tabcolsep}{2.2pt}
\resizebox{\linewidth}{!}{%
\begin{tabular}{ll ccc ccc ccc ccc ccc}
\toprule
& & \multicolumn{3}{c}{Seed 0} & \multicolumn{3}{c}{Seed 1}
  & \multicolumn{3}{c}{Seed 2} & \multicolumn{3}{c}{Seed 3}
  & \multicolumn{3}{c}{Seed 4}\\
\cmidrule(lr){3-5}\cmidrule(lr){6-8}\cmidrule(lr){9-11}\cmidrule(lr){12-14}\cmidrule(lr){15-17}
Noise & Par. & Dir. & sf & wf & Dir. & sf & wf & Dir. & sf & wf
      & Dir. & sf & wf & Dir. & sf & wf\\
\midrule
\multirow{4}{*}{Clean}
 & $\beta$   & 0.19 & 0.19 & 0.24 & 0.19 & 0.55 & 0.18 & 0.19 & 0.58 & 0.21 & 0.19 & 0.51 & 0.30 & 0.19 & 0.99 & 0.28\\
 & $\delta$  & 0.21 & 0.66 & 0.28 & 0.21 & 0.00 & 0.04 & 0.21 & 3.13 & 0.11 & 0.21 & 1.36 & 0.17 & 0.21 & 0.37 & 0.01\\
 & $\gamma$  & 0.20 & 0.10 & 0.30 & 0.20 & 0.49 & 0.26 & 0.20 & 0.62 & 0.22 & 0.20 & 0.70 & 0.40 & 0.20 & 1.09 & 0.20\\
 & $\lambda$ & 1.80 & 47.75 & 10.21 & 1.80 & 36.32 & 15.73 & 1.80 & 62.45 & 7.17 & 1.80 & 49.49 & 16.89 & 1.80 & 55.73 & 18.62\\
\midrule
\multirow{4}{*}{$4\%$}
 & $\beta$   & 2.54 & 0.45 & 0.26 & 5.45 & 0.52 & 0.31 & 0.54 & 0.57 & 0.20 & 5.56 & 0.51 & 0.24 & 2.90 & 0.37 & 0.23\\
 & $\delta$  & 5.60 & 0.08 & 0.31 & 67.80 & 0.80 & 0.14 & 30.30 & 0.96 & 0.11 & 23.88 & 0.91 & 0.24 & 7.22 & 0.04 & 0.34\\
 & $\gamma$  & 2.93 & 0.63 & 0.34 & 3.99 & 1.10 & 0.50 & 14.26 & 0.71 & 0.32 & 10.43 & 0.26 & 0.59 & 7.67 & 1.04 & 0.30\\
 & $\lambda$ & 6.99 & 34.70 & 10.93 & 8.83 & 39.23 & 21.78 & 48.03 & 70.05 & 15.60 & 29.40 & 40.57 & 19.86 & 5.87 & 43.19 & 11.83\\
\midrule
\multirow{4}{*}{$8\%$}
 & $\beta$   & 5.29 & 0.32 & 0.41 & 11.11 & 0.65 & 0.43 & 0.86 & 0.63 & 0.40 & 10.90 & 0.06 & 0.16 & 5.53 & 0.66 & 0.30\\
 & $\delta$  & 10.62 & 0.35 & 0.52 & 136.21 & 0.11 & 0.04 & 60.08 & 1.48 & 0.39 & 47.31 & 0.43 & 0.70 & 15.48 & 1.70 & 0.58\\
 & $\gamma$  & 5.96 & 0.15 & 0.30 & 8.22 & 1.35 & 0.81 & 28.20 & 0.74 & 0.38 & 20.67 & 0.75 & 0.28 & 15.63 & 0.83 & 0.17\\
 & $\lambda$ & 29.53 & 28.62 & 10.78 & 27.84 & 48.45 & 16.34 & 96.00 & 74.27 & 14.23 & 61.59 & 49.99 & 20.96 & 26.42 & 41.96 & 9.87\\
\midrule
\multirow{4}{*}{$12\%$}
 & $\beta$   & 8.02 & 0.66 & 0.46 & 16.75 & 0.76 & 0.42 & 1.27 & 0.16 & 0.40 & 16.26 & 0.24 & 0.26 & 8.13 & 1.17 & 0.41\\
 & $\delta$  & 15.91 & 0.49 & 0.46 & 204.47 & 1.62 & 0.01 & 91.36 & 0.46 & 0.69 & 71.05 & 0.11 & 1.02 & 23.98 & 2.79 & 0.78\\
 & $\gamma$  & 8.86 & 0.07 & 0.34 & 12.47 & 1.78 & 1.17 & 42.11 & 0.54 & 0.53 & 30.92 & 0.43 & 0.49 & 23.66 & 0.73 & 0.38\\
 & $\lambda$ & 49.05 & 27.40 & 12.23 & 43.52 & 40.21 & 21.36 & 100.00 & 70.11 & 15.39 & 81.98 & 51.76 & 18.25 & 44.26 & 46.79 & 6.29\\
\midrule
\multirow{4}{*}{$16\%$}
 & $\beta$   & 10.71 & 0.75 & 0.67 & 22.37 & 0.97 & 0.78 & 0.05 & 0.59 & 0.29 & 21.63 & 0.02 & 0.24 & 10.70 & 0.71 & 0.28\\
 & $\delta$  & 21.58 & 0.53 & 0.45 & 272.49 & 1.67 & 0.20 & 100.00 & 0.47 & 0.59 & 95.01 & 0.56 & 1.30 & 32.64 & 2.90 & 1.24\\
 & $\gamma$  & 11.62 & 0.00 & 0.34 & 16.73 & 2.53 & 1.59 & 54.88 & 0.48 & 0.46 & 41.18 & 0.28 & 0.76 & 31.74 & 0.83 & 0.43\\
 & $\lambda$ & 62.16 & 28.53 & 13.91 & 53.97 & 24.63 & 26.54 & 100.00 & 73.60 & 23.54 & 93.17 & 61.60 & 17.80 & 56.24 & 53.92 & 2.92\\
\midrule
\multirow{4}{*}{$20\%$}
 & $\beta$   & 13.37 & 0.87 & 0.55 & 27.95 & 0.91 & 1.26 & 2.03 & 0.73 & 0.66 & 25.55 & 0.02 & 0.02 & 13.22 & 0.22 & 0.48\\
 & $\delta$  & 27.57 & 1.13 & 1.33 & 340.26 & 2.11 & 0.06 & 100.00 & 0.04 & 0.99 & 100.00 & 0.26 & 1.89 & 41.53 & 3.94 & 2.02\\
 & $\gamma$  & 14.26 & 0.22 & 0.32 & 21.02 & 2.55 & 1.84 & 67.17 & 0.88 & 1.04 & 50.58 & 0.83 & 0.35 & 39.89 & 0.05 & 0.54\\
 & $\lambda$ & 70.70 & 27.63 & 13.92 & 60.85 & 20.94 & 31.07 & 100.00 & 73.77 & 13.87 & 99.50 & 54.24 & 24.99 & 64.06 & 42.78 & 8.91\\
\bottomrule
\end{tabular}}
\end{table}

\section{Training Dynamics}
\label{app:loss}

\begin{figure}[H]
\centering
\includegraphics[scale=0.55]{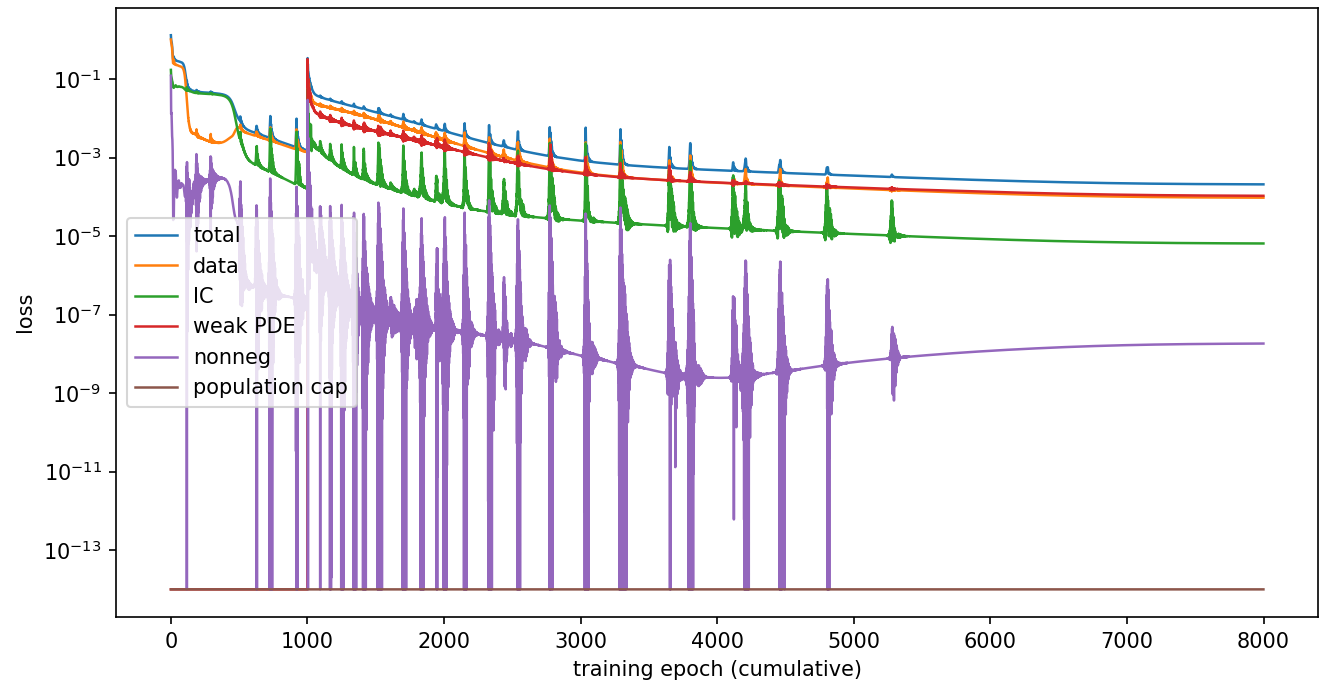}
\caption{Loss history of the sb-VPINN estimator. The weak PDE residual is active after the 1500 warmup steps of Stage A. The floor of the graph is taken as the machine zero i.e. $10^{-14}$. }
\label{fig:losshistory}
\end{figure}

Figure \ref{fig:losshistory} shows the dynamics of each loss components of the loss function \eqref{eq:total-loss} over the iterations used. During the warm-up only the data, initial condition and constraint terms are active. When the weak form PDE term is calculated, the total loss spikes and then decreases together with the data loss. The intermittent spikes, as seen in the initial condition and non-negativity terms, become smaller with increasing iterations and absent after $5300$ iterations. 

\section{Per-seed results for 2D}
\label{app:perseed_2d}

Table \ref{tab:perseed2d} is the corresponding record for the square. The table definition and format is similar to \ref{app:perseed} but the data is shown on a 2D square spatial domain.

\begin{table}[ht]
\centering
\caption{Two-dimensional square: per-seed relative parameter errors (\%) for the
direct weak inversion (Dir.), the sb-PINN (sf) and
the sb-VPINN (wf), across five seeds and six noise
levels.}
\label{tab:perseed2d}
\tiny
\setlength{\tabcolsep}{2.2pt}
\resizebox{\linewidth}{!}{%
\begin{tabular}{ll ccc ccc ccc ccc ccc}
\toprule
& & \multicolumn{3}{c}{Seed 0} & \multicolumn{3}{c}{Seed 1}
  & \multicolumn{3}{c}{Seed 2} & \multicolumn{3}{c}{Seed 3}
  & \multicolumn{3}{c}{Seed 4}\\
\cmidrule(lr){3-5}\cmidrule(lr){6-8}\cmidrule(lr){9-11}\cmidrule(lr){12-14}\cmidrule(lr){15-17}
Noise & Par. & Dir. & sf & wf & Dir. & sf & wf & Dir. & sf & wf
      & Dir. & sf & wf & Dir. & sf & wf\\
\midrule
\multirow{4}{*}{Clean}
 & $\beta$   & 0.21 & 0.49 & 0.25 & 0.21 & 0.10 & 0.27 & 0.21 & 0.30 & 0.25 & 0.21 & 0.44 & 0.16 & 0.21 & 0.36 & 0.21\\
 & $\delta$  & 0.21 & 3.20 & 0.48 & 0.21 & 4.83 & 1.17 & 0.21 & 4.16 & 0.89 & 0.21 & 3.97 & 0.39 & 0.21 & 3.39 & 0.48\\
 & $\gamma$  & 0.22 & 0.28 & 0.17 & 0.22 & 0.19 & 0.20 & 0.22 & 0.21 & 0.19 & 0.22 & 0.05 & 0.16 & 0.22 & 0.37 & 0.20\\
 & $\lambda$ & 0.48 & 87.09 & 24.12 & 0.48 & 83.87 & 14.60 & 0.48 & 45.26 & 10.06 & 0.48 & 74.71 & 22.03 & 0.48 & 74.69 & 16.72\\
\midrule
\multirow{4}{*}{$4\%$}
 & $\beta$   & 0.46 & 0.34 & 0.19 & 0.46 & 0.06 & 0.20 & 0.69 & 0.18 & 0.27 & 0.51 & 0.36 & 0.25 & 1.42 & 0.42 & 0.26\\
 & $\delta$  & 2.68 & 7.70 & 0.46 & 4.54 & 5.82 & 0.45 & 5.37 & 3.78 & 0.80 & 0.74 & 8.57 & 0.91 & 11.31 & 1.64 & 0.03\\
 & $\gamma$  & 1.38 & 0.18 & 0.19 & 1.92 & 0.51 & 0.02 & 1.31 & 0.27 & 0.14 & 1.90 & 0.06 & 0.17 & 0.79 & 0.16 & 0.32\\
 & $\lambda$ & 0.09 & 81.28 & 12.80 & 0.64 & 88.14 & 27.29 & 3.32 & 72.44 & 32.94 & 0.93 & 77.26 & 20.79 & 0.98 & 71.89 & 26.00\\
\midrule
\multirow{4}{*}{$8\%$}
 & $\beta$   & 1.16 & 0.25 & 0.05 & 1.14 & 0.03 & 0.15 & 1.61 & 0.44 & 0.32 & 0.79 & 0.44 & 0.27 & 2.61 & 0.21 & 0.31\\
 & $\delta$  & 4.76 & 5.12 & 0.06 & 9.63 & 6.85 & 0.52 & 11.32 & 5.10 & 0.72 & 2.02 & 9.38 & 1.36 & 22.11 & 3.89 & 0.20\\
 & $\gamma$  & 2.52 & 0.39 & 0.20 & 4.07 & 0.49 & 0.18 & 2.40 & 0.14 & 0.08 & 4.02 & 0.17 & 0.11 & 1.37 & 0.04 & 0.45\\
 & $\lambda$ & 2.93 & 83.07 & 15.36 & 3.62 & 90.39 & 28.94 & 9.43 & 69.62 & 35.14 & 4.24 & 78.26 & 18.98 & 4.26 & 73.87 & 28.53\\
\midrule
\multirow{4}{*}{$12\%$}
 & $\beta$   & 1.89 & 0.21 & 0.05 & 1.85 & 0.06 & 0.13 & 2.55 & 0.15 & 0.37 & 1.06 & 0.19 & 0.29 & 3.79 & 0.44 & 0.38\\
 & $\delta$  & 6.54 & 6.05 & 0.75 & 14.99 & 6.60 & 0.68 & 17.52 & 3.71 & 0.50 & 3.55 & 7.83 & 1.91 & 32.69 & 3.42 & 0.21\\
 & $\gamma$  & 3.65 & 0.48 & 0.04 & 6.23 & 0.52 & 0.32 & 3.49 & 0.57 & 0.01 & 6.14 & 0.29 & 0.07 & 1.94 & 0.29 & 0.60\\
 & $\lambda$ & 8.31 & 76.44 & 20.84 & 8.81 & 92.63 & 28.86 & 17.84 & 74.42 & 33.70 & 9.75 & 81.66 & 20.50 & 9.66 & 74.34 & 30.73\\
\midrule
\multirow{4}{*}{$16\%$}
 & $\beta$   & 2.63 & 0.18 & 0.06 & 2.57 & 0.08 & 0.05 & 3.50 & 0.25 & 0.42 & 1.31 & 0.43 & 0.25 & 4.96 & 0.36 & 0.59\\
 & $\delta$  & 8.13 & 6.80 & 1.34 & 20.49 & 6.98 & 0.50 & 23.84 & 4.96 & 0.51 & 5.23 & 9.26 & 2.36 & 43.14 & 5.53 & 0.89\\
 & $\gamma$  & 4.77 & 0.64 & 0.08 & 8.41 & 0.56 & 0.27 & 4.57 & 0.35 & 0.05 & 8.27 & 0.29 & 0.18 & 2.51 & 0.33 & 1.08\\
 & $\lambda$ & 15.29 & 83.30 & 28.32 & 15.36 & 88.69 & 22.36 & 27.45 & 72.86 & 37.60 & 16.60 & 83.42 & 14.13 & 16.37 & 87.65 & 26.58\\
\midrule
\multirow{4}{*}{$20\%$}
 & $\beta$   & 3.37 & 0.08 & 0.29 & 3.30 & 0.03 & 0.04 & 4.45 & 0.52 & 0.44 & 1.56 & 0.14 & 0.38 & 6.13 & 0.35 & 0.65\\
 & $\delta$  & 9.64 & 5.71 & 2.04 & 26.05 & 6.19 & 0.95 & 30.14 & 5.80 & 0.28 & 6.95 & 7.95 & 3.10 & 53.55 & 1.60 & 0.75\\
 & $\gamma$  & 5.88 & 0.56 & 0.31 & 10.59 & 0.59 & 0.81 & 5.65 & 0.28 & 0.16 & 10.40 & 0.43 & 0.11 & 3.09 & 0.21 & 1.16\\
 & $\lambda$ & 22.98 & 83.13 & 33.94 & 22.49 & 87.11 & 30.80 & 37.28 & 76.69 & 38.91 & 23.96 & 79.12 & 18.83 & 23.58 & 85.24 & 33.28\\
\bottomrule
\end{tabular}}
\end{table}

\end{document}